\documentclass[12pt]{article}

\usepackage{graphicx}

\usepackage{enumerate}
\usepackage{amsmath}
\usepackage{amsfonts}
\usepackage{amssymb}
\usepackage{amsfonts}
\usepackage{amssymb,amsmath,amsthm,epsfig,graphicx,natbib,subfigure,hyperref}
\usepackage[margin=2.0in]{geometry}
\usepackage{caption2}
\usepackage{color}
\usepackage{enumerate}
\usepackage{setspace}

\newtheorem{pro}{Proposition}
\newtheorem{lem}{Lemma}

\newtheorem{corr}{Corollary}
\newtheorem{ex}{Example}
\newtheorem{defin}{Definition}

\newtheorem{theorem}{Theorem}

\newtheorem{remark}[theorem]{Remark}

\newcommand {\be}{\begin{equation}}
\newcommand {\ee}{\end{equation}}
\hypersetup{colorlinks=true,linkcolor=blue,urlcolor=blue,citecolor=blue} 

\begin{document}

\title{\textbf{Structural Interventions in Networks\thanks{%
For useful comments and suggestions we thank Nizar Allouch, Francis Bloch, Yann Bramoulle,  Antonio Cabrales, George Charlson, Hanming Fang, Itay Fainmesser, Ben Golub, Sanjeev Goyal, Matthew Jackson, Ernest Liu, Evan Sadler, Adam Szeidl, Fernando Vega-Redondo, Yves Zenou, and participants at conferences and workshops.} }}
\author{Yang Sun\thanks{%
Department of Economics, Sichuan University, China. \textit{Email}:
sunyang789987@gmail.com} \and Wei Zhao\thanks{%
Department of Economics and Decision Sciences, HEC Paris, France. \textit{%
Email}: wei.zhao1@hec.edu } \and Junjie Zhou\thanks{%
Department of Economics, National University of Singapore, Singapore. 
\textit{Email}: zhoujj03001@gmail.com} }
\date{\today}
\maketitle

\begin{abstract}
Two types of interventions are commonly implemented in networks:
characteristic intervention, which influences individuals' intrinsic incentives, and structural intervention,
which targets  the social links among individuals. 
 In this paper we provide a general framework to evaluate the distinct equilibrium effects of
both types of interventions. We identify a hidden equivalence between a
structural intervention and an \emph{endogenously determined} characteristic
intervention. Compared with existing approaches in the literature, the
perspective from such an equivalence provides several advantages 
in the analysis of interventions that target  network
structure. We present a wide range of applications of our theory, including
identifying the most wanted criminal(s) in delinquent networks and targeting
the key connector for isolated communities. 
\end{abstract}

\setlength{\textwidth}{15cm} \setlength{\textheight}{21.5cm}

\newpage

\section{Introduction}

\label{sec:intro}

Social ties shape economic agents' decisions in a connected world, ranging
from which product to buy for consumers, how much time to spend  studying
for pupils, how much effort to exert for workers on a team, whether to
commit a crime for teenagers, etc.\footnote{%
Numerous studies have highlighted the influence of networks in different
contexts such as microfinance (\citet{banerjee2013diffusion}); firm
performance (\cite{Cai2017}); productivity at work (\cite{mas2009peers});
R\&D (\cite{goyal2001RD}); education (\cite%
{sacerdote2001peer,Calvo-Armengol2009}); crime (\cite{Ballester2006});
public goods provision (\citet{BRAMOULLE2007478,Allouch2015}); brand choice (%
\cite{Godes2004}); and policy intervention \citet{Galeotti2017}). For recent
surveys, see, for instance, %
\citet{bramoulle2016oxford,jackson2017economic,NEP2019}.} These social ties,
structurally represented as a network, govern individual incentives and %
therefore collectively determine equilibrium outcomes and welfare in the society.  
Thus, structural intervention in social ties  provides an important policy instrument for the social planner. 
A natural research question arises: how to best intervene in the social structure to maximize a certain performance
 objective subject to  certain resource constraints. This research problem is
inherently difficult, as it is well known that networks operate in a complex
manner. Local changes in social links between a few nodes can influence the actions
of a large set of nodes, including those that are far away 
through ripple effects. Furthermore, the influence is not homogeneous: Nodes
that are closer to (further away from) the origin of shocks tend to be more (less) responsive. These key features of shock
propagation and heterogeneous responses make the analysis of structural
interventions both intriguing and challenging.\footnote{%
Admittedly, these two features are also true for other types of
intervention, such as the characteristic intervention. As shown in Section %
\ref{sec:2.2}, the problem of characteristic intervention in a fixed network
is much simpler and has been extensively studied in the literature.}

In this paper, we propose a general yet tractable framework to
quantitatively assess the consequences of an arbitrary structural
intervention in social ties on  equilibrium actions. By overcoming the
challenges described above, we present a neat characterization result in
Proposition \ref{prop-1}  that evaluates the change in equilibrium behavior
in response to changes in the network structure. We then apply Proposition %
\ref{prop-1} to several economic settings, such as key group removal in
delinquent networks (in Section \ref{s-key group})
 and key connectors for isolated communities (in Section \ref{s-Bridge}).

More specifically, our model of structural intervention builds on a seminal
paper by \citet{Ballester2006} (BCZ hereafter), who propose a simple yet
powerful model of interactions in a fixed network.\footnote{%
The  model in BCZ has been applied, empirically tested, and generalized extensively
 in the network literature; see, for
example, \citet{Calvo-Armengol2009,D2x,Galeotti2017}.} They identify the
equivalence between equilibrium actions in a network game and the
Katz-Bonacich centralities in sociology (\cite{bonacich1987}). The
Katz-Bonacich centrality of a node on a network simply counts the sum of
geometrically discounted walks originating from this node to all other
nodes in the network, weighted by the characteristics of  the ending nodes.\footnote{%
This Katz-Bonacich centrality (and its variants and generalizations) plays
important roles in shaping agents' decisions in a wide range of network
models; see, for instance, production networks (\citet{Acemoglu2012,baqaee2018cascading, liu2019industrial}) and
the pricing of social products (\citet{candogan2012optimal, BLOCH2013243,D2p}).} 
 Proposition \ref{prop-1} in our paper characterizes the impacts of
structural intervention on the equilibrium by employing another equivalence
result: Any (local) intervention on the network structure is
equivalent to a (local) \emph{endogenously determined} intervention on characteristics.
Specifically, we find that the equilibrium induced by a structural
intervention coincides with that induced by  an endogenously
determined characteristic intervention without changing the network
structure. Moreover, the endogenously determined characteristic intervention
only changes the characteristics of the players whose social ties are
altered by the structural intervention. The analysis of 
post-intervention equilibrium becomes much simpler after
translating a structural intervention to the characteristic intervention,
since the  latter changes an individual's equilibrium behavior
linearly and is well studied in the literature, while the former is nonlinear.
 Furthermore, in Corollary \ref{cor-1}, we provide a sufficient condition on  a structural intervention
  to induce higher aggregate equilibrium activity, and use it to
   check the effect of a link reallocation or a link swap on the aggregate action.

For applications, we first adopt the outcome equivalence result to study 
the  key group problem, which aims to specify the group of players, that if removed,
 reduces the aggregate equilibrium activity the most.
Specifically, the removal of a group of players is equivalent to a certain
characteristic intervention restricted to this group. Such a characteristic
intervention is chosen to ensure that within the original network, 
 the  induced equilibrium efforts of nodes in this group reduce to
zero. As a generalization of  the single node intercentrality given by BCZ, we
provide a closed-form index of group inter-centrality explicitly. Such an index 
takes into account both the  activities  nodes in this group and their influences on
nodes outside  the group. The group intercentrality index
reveals that the higher the connectedness between nodes within a group, the
lower the intercentrality of the group. Therefore, the greedy algorithm,
which sequentially selects nodes with the highest single node intercentrality, may fail to
find the key group with the highest intercentrality. We also show that the group
intercentrality index is equivalent to the aggregate  sum
of all walks that must pass the group. As a by-product, we
characterize the aggregate sum of all walks starting from
one group and ending at the second group, which does \emph{not} pass the third
group. This result generalizes some of the findings on
the targeting centrality proposed by \cite{Bramoulle2018}
in an information diffusion setting.

Next, we introduce a bridge index to characterize the impact of building 
a bridge between separated networks (Proposition \ref{pro-bridge}). We use
the bridge index to fully solve the key bridge problem. Furthermore, we show
that the key bridge player must locate at the Pareto frontier of
Katz-Bonacich centrality and self-loop in the network.  In general, the
selection of a bridge pair is an interdependent decision across two networks, since
 the identity of a key bridge player in one network depends on who is
selected as his partner in the second network. These findings are summarized
in Corollary \ref{bridge cor} and illustrated in Example \ref{ex-3}. We also
extend the analysis to consider the value of an existing link (the key link
problem) and the value of a potential link for an arbitrary network in
Section \ref{sec:4.2}.  As an illustration, we compare intergroup links and intragroup links
 in Example \ref{ex-4}.

Our paper builds  on the vast literature on network games
(see \cite{Ballester2006}; \cite{BRAMOULLE2007478}; and \cite{Galeotti2010}).
These papers typically characterize the effects of network structure on
equilibrium behavior. Our paper instead focuses on how interventions on
 network structures affect equilibrium outcomes and sheds light on 
 policy design that targets at network structure.

The literature on interventions in networks can be broadly divided into two
categories: characteristic intervention and network structure intervention. 
In the first category, the characteristics of individuals can be changed by subsidy
or taxation on  choices. For instance, \cite{DEMANGE2017} and \citet{Galeotti2017} study the optimal intervention on characteristics subject to a fixed budget constraint and a quadratic adjustment cost, respectively.
  Motivated by \cite{Ballester2006, Ballester2010}, our paper mainly focuses on the second category:
structural intervention. The identified equivalence  between a structural intervention and an endogenously determined characteristic intervention in our paper provides an interesting link between these two categories.  Several papers on network formation study the most efficient network by analyzing the impact of link shifting (e.g., \cite{Belhaj2016} and \cite{Li2019}). As a complementary result, we propose a sufficient condition to guarantee that a structural intervention leads to higher aggregate action.

An important topic in social networks is  the relative importance of
a node in a given network using diverse indices. Various centrality measures have
been proposed to serve the purpose. \cite{bloch2016centrality} take an
axiomatic approach to provide a unified perspective on several commonly used
centrality measures. \cite{Ballester2006} give a micro-foundation of
Katz-Bonacich centrality and propose another measure -- i.e., intercentrality --
to characterize the impact of a node removal. Analogous measures for a group
of nodes, instead of a single node, are not fully developed. One exception
is \cite{Ballester2010}, who define  group intercentrality. One of our
contributions is proposing a specific form of group intercentrality  using
the statistics in the underlying network; we show that 
 group intercentrality decreases with connectedness between
group members. \cite{Bramoulle2018} study the contribution of a pair of nodes (one sender and one receiver) in
an information transmission setting.   Our analysis in Section \ref{sec:3.2}
can be viewed as an extension of their results by allowing 
multiple senders and multiple receivers.

This paper also speaks to the literature  on the effect of
bridge(s) between isolated communities. \cite{Cai2017} demonstrate that business
meetings facilitate interfirm communications and create enormous economic
value by increasing firm performance. To the best of our knowledge, \cite%
{Golub2010} is the only network paper to theoretically study the impact of
bridge. They consider a social learning model, and  their main
focus is on the eigenvalue centrality. Our paper, instead, studies the impact
of a bridge on Katz-Bonacich centralities and proposes an explicit  bridge
index to characterize the key pair of nodes connecting two separated networks. See 
\cite{Golub2010} for a comprehensive discussion of the related literature.



\section{Interventions in networks: Theory}

\label{The model}

\subsection{Setup}

\textsc{Baseline game played on a network}~ Consider a network game played
by a set of players $N=\left\{ 1,2,\ldots ,n\right\} $ embedded in a social
network $\mathbf{G}$, which is represented by an $n\times n$ adjacency matrix 
$\mathbf{G}=\left( g_{ij}\right) _{n\times n}$. Each player $i$ chooses an
effort $a_{i}\in \lbrack 0,\infty)$ simultaneously with %
 payoff function given as follows:\footnote{%
In Section \ref{extensions}, we discuss several extensions of the baseline
model.} 
\begin{equation}
u_{i}\left( a_{i},\mathbf{a}_{-i}\right) =\theta _{i}a_{i}-\frac{1}{2}%
a_{i}^{2}+\delta \sum_{k=1}^{n} g_{ik}a_{i}a_{k},~~~~i\in
N.  \label{eq:complements}
\end{equation}%
This specification of payoff closely follows from \citet{Ballester2006},
where $\theta _{i}$ measures player $i$'s intrinsic marginal utility (hence $%
i$'s characteristic), $\frac{1}{2}a_{i}^{2}$ denotes player $i$'s cost of
effort, and the last term, $\delta \underset{k=1}{\overset{n}{\sum }}%
g_{ik}a_{i}a_{k}$, captures the interaction term that represents local
network effects among players. The scalar parameter $\delta $ controls the
strength of network interaction. We assume $\delta >0$ so the game exhibits
strategic complementarity. We use $\Gamma \left( \mathbf{G},\boldsymbol{%
\theta },\delta \right) $ to denote the network game represented above,
where $\boldsymbol{\theta }=\left( \theta _{1},\ldots ,\theta _{n}\right)
^{\prime }$ is the characteristics vector.

Throughout the paper, we impose the standard assumptions
that (i) $\mathbf{G} $ is symmetric with $g_{ij}=g_{ji}\in \left\{
0,1\right\} $, and (ii) $g_{ii}=0$ for all $i\in N$.\footnote{%
For ease of interpretation, we focus on undirected zero-one network matrix $%
\mathbf{G}$. Our results can be easily generalized to weighted directed
networks.} Let $\lambda _{\max }\left( \mathbf{G}\right) $ denote the
spectral radius of matrix $\mathbf{G}$. By Perron-Frobenius theorem, $%
\lambda _{\max }\left( \mathbf{G}\right) $ also equals the largest
eigenvalue of $\mathbf{G}$. The following is a well-known measure of
centralities in network literature.

\begin{defin}
\label{def-KB} Given a network $\mathbf{G}$, a scalar $\delta $, and an $n$%
-dimensional vector $\boldsymbol{\theta }$, we define $\boldsymbol{\theta }$%
-weighted Katz-Bonacich centralities as 
\begin{equation}
\mathbf{b}\left( \mathbf{G},\boldsymbol{\theta },\delta \right)
=(b_{1},b_{2},\cdots ,b_{n})^{\prime }\equiv \left( \mathbf{I}-\delta \mathbf{G}%
\right) ^{-1}\boldsymbol{\theta },
\end{equation}%
provided that $\delta <\frac{1}{\lambda _{\max }\left( \mathbf{G}\right) }$.
When $\boldsymbol{\theta }=(1,1,\cdots ,1)^{\prime }=
\mathbf{1}$, we call $\mathbf{b}\left( \mathbf{G},\delta \right) \equiv 
\mathbf{b}\left( \mathbf{G},\boldsymbol{1},\delta \right) $ the unweighted
Katz-Bonacich centralities. Define the Leontief inverse matrix 
\begin{equation}
\mathbf{M}\left( \mathbf{G},\delta \right) =\left( m_{ij}\left( \mathbf{G}%
\right) \right) _{n\times n}\equiv \left( \mathbf{I}-\delta \mathbf{G}%
\right) ^{-1},
\end{equation}%
so that $b_{i}\left( \mathbf{G},\boldsymbol{\theta },\delta \right) =%
\underset{j=1}{\overset{n}{\sum }}m_{ij}\left( \mathbf{G}\right) \theta _{j}$%
.
\end{defin}

Intuitively, $m_{ij}\left( \mathbf{G}\right) $ counts the total number of
walks from $i$ to $j$ in network $\mathbf{G}$ with path of
length $k$ discounted by $\delta ^{k}$. So $i$'s Katz-Bonacich centrality $%
b_{i}\left( \mathbf{G},\boldsymbol{\theta },\delta \right) $ is 
the sum of walks starting from $i$ and ending at any node $j$ with weights $%
\theta _{j}$.\footnote{%
This follows from the following identity of the Leontief inverse matrix: 
\begin{equation*}
\mathbf{M}\left( \mathbf{G},\delta \right) =\left( \mathbf{I}-\delta \mathbf{%
G}\right) ^{-1}=\mathbf{I}+\delta \mathbf{G}+\delta ^{2}\mathbf{G}%
^{2}+\cdots .
\end{equation*}%
This Neumann series converges when $0\leq \delta <\frac{1}{\lambda _{\max
}\left( \mathbf{G}\right) }$.} 
Interestingly, \citet{Ballester2006} show
that when $\delta <\frac{1}{\lambda _{\max }\left( \mathbf{G}\right) }$,
game $\Gamma \left( \mathbf{G},\boldsymbol{\theta },\delta \right) $ has a
unique Nash equilibrium in which each player $i$'s equilibrium action $x_{i}^{\ast }$ is exactly equal to $i$'s Katz-Bonacich centrality, i.e., 
\begin{equation}
\mathbf{x}^{\ast }=\mathbf{b}\left( \mathbf{G},\boldsymbol{\theta },\delta
\right) \text{.}  \label{eq:BCZ06}
\end{equation}
More influential players, measured by  Katz-Bonacich
centralities, are more active in equilibrium. Such an elegant relationship
between equilibrium outcomes and Katz-Bonacich centralities is the starting
point of our analysis.

\noindent \textsc{Interventions on networks}~ The network structure $\mathbf{%
G}$ and the characteristics vector $\boldsymbol{\theta }$ jointly shape the
equilibrium actions of players and welfare in $\Gamma \left( \mathbf{G},%
\boldsymbol{\theta },\delta \right) $. We introduce two primary types of
interventions to influence the equilibrium outcomes: \emph{characteristic
intervention} and \emph{structural intervention}. In the former case $%
\boldsymbol{\theta }$ is modified to $\boldsymbol{\hat{\theta}}$  and $%
\mathbf{G}$ is fixed, and in the latter case $\mathbf{G}$
is changed to $\mathbf{\hat{G}}$  and $\boldsymbol{\theta }$ is fixed. The
economic consequences of these two types of interventions are characterized
in detail in the next two subsections. The hybrid case involving both types of
interventions is discussed in Section \ref{sec:hybrid}. Importantly, changes 
in either $\mathbf{G}$ or $\boldsymbol{\theta }$ in our
paper occur for exogenous and independent reasons. Furthermore, we do not
consider the possibility that changes in characteristics can induce changes
in the network structure, and vice versa. The parameter $\delta $ is fixed
throughout the paper, and is often omitted in  expressions when the context
is clear.

\noindent \textsc{Assumptions}~ To ensure the uniqueness of Nash equilibrium
before and after the intervention, we impose the standard spectral
condition: $\delta <\frac{1}{\lambda _{\max }\left(\mathbf{G}\right) }$ and $%
\delta <\frac{1}{\lambda _{\max }\left(\mathbf{\hat G}\right) }$. Our main
focus is on the effects of interventions on equilibrium actions. In
applications, we  analyze optimal intervention under 
certain resource constraints (such as limiting the number of links or
players that can be intervened). Beyond that, we do not explicitly model the
cost side of interventions.\footnote{%
With parametric assumptions on the cost of interventions, certainly more can
be said about optimal interventions for the social planer.
Assuming  a quadratic loss function of the Euclidean distance between $%
\boldsymbol{\theta}$ and $\boldsymbol{\hat\theta}$, \citet{Galeotti2017}
explicitly solve the optimal characteristic intervention, for both the case
with $\delta>0$ (strategic complement) and that with $\delta<0$ (strategic
substitute). They relate the optimal interventions to the spectral 
properties of interaction networks and show that the optimal intervention
takes a simple form when the planner's budget is sufficiently large.}

\noindent \textsc{Notation} Before proceeding, we introduce some notation.
 In the network $\left( N,\mathbf{G}\right) $,
for any subset $A\subseteq N$, we let $|A|$ denote the cardinality of this
set, and let $A^{C}=N\backslash A$ denote the complement of $A$. Let $%
\mathbf{G}_{AA}$ denote the $|A|\times |A|$ adjacency matrix of the
subnetwork formed by players in $A$. Moreover, the adjacency matrix $\mathbf{%
G}$ can be written as a block matrix $\mathbf{G}=%
\begin{bmatrix}
\mathbf{G}_{A^{C}A^{C}} & \mathbf{G}_{A^{C}A} \\ 
\mathbf{G}_{AA^{C}} & \mathbf{G}_{AA}%
\end{bmatrix}%
.$ Similarly, we can rewrite a column vector $\mathbf{x}$ of length $n$ as $%
\begin{bmatrix}
\mathbf{x}_{A^{C}} \\ 
\mathbf{x}_{A}%
\end{bmatrix}%
$. We use $x$ to denote the sum of all elements 
in vector $\mathbf{x=}\left(x_{i}\right) _{n\times 1}$, i.e., $x=\underset{i=1}{\overset{n}{\sum }}%
x_{i} $. The transpose of a matrix $\mathbf{H}$ is denoted by $\mathbf{H}%
^{\prime } $. Consider two matrices $\mathbf{Q=}\left( q_{ij}\right)
_{n\times m}$ and $\mathbf{P}=\left( p_{ij}\right) _{n\times m}$ of the same
dimension. We write $\mathbf{Q}\succeq (\preceq )\mathbf{P}$ if and only if $%
q_{ij}\geq (\leq ){p}_{ij}$ for any $i$, $j$.

\subsection{Effects of a characteristic intervention}

\label{sec:2.2}

In this subsection, we consider the impact of characteristic intervention.
In reality, the characteristics of players can be increased by subsidy or
decreased by taxation (see, for example, \citet{Galeotti2017} for further
illustrations of changing $\boldsymbol{\theta }$). 
The characteristic intervention changes the characteristic vector
 $\boldsymbol{\theta }$ to $\boldsymbol{\hat{\theta}}$. 
 The game after intervention $\Gamma \left( \mathbf{G},\boldsymbol{\hat{%
\theta}}\right) $ reaches a new equilibrium, denoted as $\mathbf{\hat{x}}%
^{\ast }$. Define $\Delta \boldsymbol{\theta }:=\boldsymbol{\hat{\theta}}-%
\boldsymbol{\theta }$ as the differences in players' characteristics and $%
\Delta \mathbf{x}^{\ast }=\mathbf{\hat{x}}^{\ast }-\mathbf{x}^{\ast }$ as
the changes in equilibrium actions. Since interventions can be targeted, not
every player is equally affected; thus $\Delta \theta _{i}$ may not have the
same sign or magnitude as $\Delta \theta _{j}$. Define $S=\left\{ i\in
N:\Delta \theta _{i}\neq 0\right\} $ as the set of players involved in this
characteristic intervention, and we rewrite $\Delta \boldsymbol{\theta }$ as $%
\begin{bmatrix}
\mathbf{0} \\ 
\Delta \boldsymbol{\theta }_{S}%
\end{bmatrix}%
$ after suitable relabelling of players. That is,  the characteristics of players
in $S$ are changed by $\Delta \boldsymbol{\theta }_{S}$, while the
characteristics of players in its complement $S^{C}$ are not affected. The
following Lemma summarizes the effects of a characteristic intervention.

\begin{lem}
\label{Partial} \label{lm-1} After characteristic intervention $\Delta 
\boldsymbol{\theta}=%
\begin{bmatrix}
\mathbf{0} \\ 
\Delta \boldsymbol{\theta }_{S}%
\end{bmatrix}$, the change in equilibrium  is 
\begin{equation}
\Delta \mathbf{x}_{A}^{\ast }=\mathbf{M}_{AS}\left( \mathbf{G}\right) \Delta 
\boldsymbol{\theta }_{S}  \label{eq:partial}
\end{equation}%
for any subset $A\subseteq N$. Moreover, the change in the aggregate action
is 
\begin{equation}
\Delta x^{\ast }= \sum_{i\in N} \Delta x_i^*=\mathbf{b}_{S}^{\prime }\left( 
\mathbf{G}\right) \Delta \boldsymbol{\theta }_{S}\text{.}
\label{eq:aggregate partial}
\end{equation}
\end{lem}

This Lemma is  straightforward according to equation \eqref{eq:BCZ06},
 since the network structure is fixed during the intervention and the equilibrium action profile $\mathbf{x}^{\ast }$ is linear in the characteristics vector $\boldsymbol{\theta }$ with sensitivity matrix given by $\mathbf{M}(\mathbf{G})$. 
In particular, consider a characteristic intervention at a single node $%
S=\{j\} $ by $\Delta \theta _{j}$; then for $A=\{i\}$, $\Delta x_{i}^{\ast
}=m_{ij}\left( \mathbf{G}\right) \Delta \theta _{j}$ by Lemma \ref{Partial}.
The marginal contribution of $j$'s characteristics on $i$'s equilibrium
behavior is exactly $m_{ij}(\mathbf{G})$: the total number of walks from $i$
to $j$ with length discount $\delta$ in the network.
Summing over all $i$, the marginal contribution of $j$'s characteristics on
the aggregate effort is just $\sum_{i\in N}m_{ij}(\mathbf{G})=\sum_{i\in
N}m_{ji}(\mathbf{G})=b_{j}\left( \mathbf{G}\right) $.\footnote{%
We exploit the symmetry of matrix $\mathbf{M}$ here.} In other words, we
have 
\begin{equation}
\frac{\partial x^{\ast }}{\partial {\theta }_{j}}=\frac{\partial
\{\sum_{i\in N}x_{i}^{\ast }\}}{\partial {\theta }_{j}}=b_{j}(\mathbf{G}).
\label{eq:x-i}
\end{equation}%
When the characteristics of multiple players are modified during the
intervention (so $S$ contains multiple players), by Lemma \ref{Partial}, we
observe a form of linearity: The change in player $i$'s equilibrium action
is simply the sum, over $j$ in $S$, of the effect caused by $j$ i.e., $\Delta
x_{i}^{\ast }=\sum_{j\in S}m_{ij}\left( \mathbf{G}\right) \Delta \theta _{j}$%
. In other words, 
\begin{equation}
\frac{\partial \mathbf{x}_{A}^{\ast }}{\partial \boldsymbol{\theta }_{S}}=%
\mathbf{M}_{AS}\left( \mathbf{G}\right) 
\mbox{for any subset $A\subseteq
N$.}  \label{eq:xAS}
\end{equation}%
As we will see in the next subsection, this desirable feature of linearity does
not hold for structural intervention, the effects of which are nonlinear and
hence more complex to analyze.

\subsection{Effects of a structural intervention}

\label{sec:2.3}

In this subsection, we study the impact of structural intervention, i.e.,
changing $\mathbf{G}$ to $\mathbf{\hat G}$. The equilibrium action profile
changes from $\mathbf{x}^*$ in the original game $\Gamma \left( \mathbf{G},%
\boldsymbol{\theta},\delta\right)$ to $\mathbf{\hat x}^*$ in the new game $%
\Gamma \left( \mathbf{\hat G},\boldsymbol{\theta},\delta\right)$. Define $%
\mathbf{C}=\mathbf{\hat{G}} - \mathbf{G}$ as the change in the network
structure and $\Delta\mathbf{\hat x}^*=\mathbf{\hat{x}}^{\ast }-\mathbf{x}%
^{\ast }$ as the change in equilibrium actions. Structural interventions may
occur when new links are formed and/or existing links or nodes are deleted.
The matrix $\mathbf{C}=\left(c_{ij}\right) _{n\times n}$ is symmetric with
entries in $\left\{1, 0, -1\right\}$. In particular, $\mathbf{C}$ is not
necessarily a nonnegative matrix. Let $S=\left\{ i\in N:c_{ij}\neq 0\text{
for some }j\in N\right\} $ denote the set of players involved in this
intervention. Rearranging the order of players if necessary, we can
represent the intervention matrix $\mathbf{C}$ by the  block matrix 
$%
\begin{bmatrix}
\mathbf{0} & \mathbf{0} \\ 
\mathbf{0} & \mathbf{C}_{SS}%
\end{bmatrix}%
.$

To state the next result regarding the effects of the structural
intervention $\mathbf{C}$, we define an $|S|$-dimensional
vector $\boldsymbol{\Delta \theta }_{S}^{\ast }$ as follows: 
\begin{equation}
\boldsymbol{\Delta \theta }_{S}^{\ast }\equiv \delta \mathbf{C}_{SS}\left( 
\mathbf{I}-\delta \mathbf{M}_{SS}\left( \mathbf{G}\right) \mathbf{C}%
_{SS}\right) ^{-1}\mathbf{b}_{S}\left( \mathbf{G},\boldsymbol{\theta }%
\right) .  \label{eq:weight intervention}
\end{equation}%
Note that this vector can  be computed easily using centralities measures
before the intervention (such as $\mathbf{b}_{S}\left( \mathbf{G}\right) $
and $\mathbf{M}_{SS}\left( \mathbf{G}\right) $), and the intervention matrix 
$\mathbf{C}_{SS}$.

\begin{lem}[Equivalence between structural intervention and characteristics
intervention]
\label{lm-2} Start with $\Gamma \left( \mathbf{G},\boldsymbol{\theta}%
,\delta\right)$. A structural intervention $\mathbf{C}=%
\begin{bmatrix}
\mathbf{0} & \mathbf{0} \\ 
\mathbf{0} & \mathbf{C}_{SS}%
\end{bmatrix}%
$ has the same effects on equilibrium actions as a characteristics
intervention $\widetilde{\Delta \boldsymbol{\theta}}\equiv 
\begin{bmatrix}
\mathbf{0} \\ 
\boldsymbol{\Delta \theta }_{S}^{\ast }%
\end{bmatrix}%
,$ where $\boldsymbol{\Delta \theta }_{S}^{\ast}$ is given in equation \eqref{eq:weight intervention}.
\end{lem}

The main idea behind Lemma \ref{lm-2} is very simple. An equilibrium is a
fixed point of the best-response mapping, which, in the framework of BCZ, is
linearly additively separable in actions $\mathbf{x}$ and characteristics $%
\boldsymbol{\theta }$. That is, $\mathbf{x}^{\ast }$ is an
equilibrium of game $\Gamma \left( \mathbf{G},\boldsymbol{\theta },\delta
\right) $ if and only if $\mathbf{x}^{\ast }=\boldsymbol{\theta }+\delta 
\mathbf{G}\mathbf{x}^{\ast }$. We could reinterpret the post-intervention
equilibrium $\mathbf{\hat{x}}^{\ast }$ as a fixed point in the
pre-intervention game after modifying the characteristics vector of players
in $S$ from $\boldsymbol{\theta }_{S}$ to $\boldsymbol{\theta }_{S}+%
\boldsymbol{\Delta \theta }_{S}^{\ast }$ with $\boldsymbol{\Delta \theta }%
_{S}^{\ast }=\delta \mathbf{C}_{SS}\mathbf{\hat{x}}_{S}^{\ast }$.\footnote{%
Formally, the post-intervention equilibrium action profile $\mathbf{\hat{x}}$
solves 
\begin{eqnarray*}
\mathbf{\hat{x}}^{\ast } &=&\boldsymbol{\theta }+\delta \left( \mathbf{G}+%
\mathbf{C}\right) \mathbf{\hat{x}}^{\ast }=\left( \boldsymbol{\theta }%
+\delta \mathbf{C}\mathbf{\hat{x}}^{\ast }\right) +\delta \mathbf{G\hat{x}}%
^{\ast }=\left( \boldsymbol{\theta }+%
\begin{bmatrix}
\mathbf{0} \\ 
\delta \mathbf{C}_{SS}\mathbf{\hat{x}}_{S}^{*}%
\end{bmatrix}%
\right) +\delta \mathbf{G\hat{x}}^{\ast }.
\end{eqnarray*}%
Put differently, we have $\mathbf{\hat{x}}^{\ast }=\mathbf{b}(\mathbf{G},%
\boldsymbol{\theta }+\widetilde{\Delta \boldsymbol{\theta }})$ with $%
\widetilde{\Delta \boldsymbol{\theta }}=%
\begin{bmatrix}
\mathbf{0} \\ 
\delta \mathbf{C}_{SS}\mathbf{\hat{x}}_{S}^{*}%
\end{bmatrix}%
$.} To determine $\mathbf{\hat{x}}_{S}^{\ast }$, which is endogenous, we
make use of the following identity: 
\begin{equation*}
\underbrace{\mathbf{\hat{x}}_{S}^{\ast }-\mathbf{b}_{S}}_{=\Delta \mathbf{x}%
_{S}^{\ast }}=\mathbf{M}_{SS}\underbrace{\left( \delta \mathbf{C}_{SS}%
\mathbf{\hat{x}}_{S}^{\ast }\right) }_{=\boldsymbol{\Delta \theta }%
_{S}^{\ast }}.\footnote{%
Solving this yields $\mathbf{\hat{x}}^{\ast }=\left( \mathbf{I}-\delta \mathbf{%
M}_{SS}\left( \mathbf{G}\right) \mathbf{C}_{SS}\right) ^{-1}\mathbf{b}%
_{S}\left( \mathbf{G},\boldsymbol{\theta }\right) $, which leads to $%
\boldsymbol{\Delta \theta }_{S}^{\ast }$, as given in equation \eqref{eq:weight
intervention}.}
\end{equation*}%
The above identity follows from equation \eqref{eq:partial}: The
term on the left-hand side is just the change in the equilibrium profile of $S$ (recall 
that $\mathbf{b}_{S}$ is just the pre-intervention effort profile of $%
S $), and the term on the right-hand side follows from the above equivalent
characteristic reinterpretation of structural intervention.

Lemma \ref{lm-2} demonstrates a simple equivalence between a structural
intervention and an \emph{endogenously determined} characteristics
intervention. Lemma \ref{lm-2}, combined with Lemma \ref{lm-1}, greatly
simplifies  analysis of the effects of structural interventions in networks. 
Three key features are worth noting. The first is \emph{locality}. The vector of
the equivalent characteristics intervention is nonzero only on $S$ (the set
of nodes involved in the intervention) and it only requires information on $%
\mathbf{M(G)}$ and $\mathbf{b(G)}$ of nodes in $S$, i.e., the entries of $%
\mathbf{M}_{SS}(\mathbf{G})$ and $\mathbf{b}_{S}(\mathbf{G})$. 
This feature is appealing, since in many applications $|S| $ is relatively 
small compared with network
size $|N|$ (see our applications in subsequent sections). Locality makes 
expression of the effects of structural interventions much more succinct, and
hence easier to interpret.

The second feature is \emph{convenience}. In addition to the intervention $\mathbf{C}$,
 determination of the equivalent characteristics intervention uses the
Leontief inverse matrix $\mathbf{M}(\mathbf{G})$ and Katz-Bonacich
centralities $\mathbf{b}(\mathbf{G},\boldsymbol{\theta })$ evaluated \emph{%
before} the intervention,  rather than the indices of
post-intervention network $\hat{\mathbf{G}}$. Since this information on
pre-intervention centralities is usually available,  the amount of additional information 
needed to evaluate structural intervention is minimal. The second feature also renders
the comparative analysis across different structural interventions
manageable. To compare the effects of two structural interventions, say $%
\mathbf{C^{\prime }}$ and $\mathbf{C^{\prime \prime }}$, we keep track of
the differences in the vectors of characteristics interventions by mainly
focusing on the differences between $\mathbf{C^{\prime }}$ and $\mathbf{%
C^{\prime \prime }}$, since information on  centralities comes from a common
source: the pre-intervention equilibrium.

The third key feature is \emph{simplicity.} The characteristic intervention affects players' equilibrium efforts
linearly with sensitivity matrix $\mathbf{M}\left( \mathbf{G}%
\right) $ (see Lemma \ref{lm-1}). In contrast, the impact of structural
intervention is much more involved. Using the Newmann series definition (or
the walk-counting explanations) of centrality measures, we obtain the
following decomposition of the changes in actions: 
\begin{eqnarray*}
\Delta \mathbf{x}^{\ast } &=&\left( \mathbf{I}-\delta \left( \mathbf{G+C}%
\right) \right) ^{-1}\boldsymbol{\theta }-\left( \mathbf{I}-\delta \mathbf{G}%
\right) ^{-1}\boldsymbol{\theta } \\
&=&\left\{ \delta \mathbf{C}+\delta ^{2}\left( \left( \mathbf{G+C}\right)
^{2}-\mathbf{G}^{2}\right) +\delta ^{3}\left( \left( \mathbf{G+C}\right)
^{3}-\mathbf{G}^{3}\right) +\cdots \right\} \boldsymbol{\theta }\text{.}
\end{eqnarray*}%
For each $k=1,2,\cdots $, the term $\left( \mathbf{G+C}\right) ^{k}-\mathbf{G%
}^{k}$ keeps track of changes in the number of walks with length $k$ due to
this intervention $\mathbf{C}$. Evaluating this term directly is
increasingly complicated as $k$ gets larger.  By transforming the
structural intervention to an \emph{endogenously determined} characteristic intervention,
Proposition \ref{prop-1} bypasses most of the challenging issues associated
with the evaluation of structural interventions.

The following Proposition immediately follows from Lemmas \ref{lm-1} and \ref%
{lm-2}.

\begin{pro}[Effects of structural interventions]
\label{prop-1} After structural intervention $\mathbf{C}=%
\begin{bmatrix}
\mathbf{0} & \mathbf{0} \\ 
\mathbf{0} & \mathbf{C}_{SS}%
\end{bmatrix}%
$,

\begin{enumerate}
\item[(i)] the change in equilibrium  is 
\begin{equation}
\Delta \mathbf{x}_{A}^{\ast }=\mathbf{M}_{AS}\left( \mathbf{G}\right) \Delta 
\boldsymbol{\theta }_{S}^{\ast }
\end{equation}%
for any subset $A\subseteq N$; and

\item[(ii)] the change in the aggregate action is 
\begin{equation}
\Delta x^{\ast }=\mathbf{b}_{S}^{\prime }\left( \mathbf{G}\right) \Delta 
\boldsymbol{\theta }_{S}^{\ast },  \label{eq-prop1-x}
\end{equation}%
where $\boldsymbol{\Delta \theta }_{S}^{\ast }$ is given in equation
\eqref{eq:weight
intervention}.
\end{enumerate}
\end{pro}

Proposition \ref{prop-1} is applicable to an arbitrary structural
intervention. In what follows, we present several examples of structural
interventions that are commonly used in the network literature, though
under different contexts.

\begin{ex}[Different types of structural interventions] \label{ex-1} ~~~

\begin{enumerate}
\item[(i)] Creating a new link between $i$ and $j$: $\mathbf{C}=\mathbf{E}%
_{ij}$.\footnote{%
For instance, \cite{Golub2010} evaluate the impact of adding a link (a weak
tie) between two disconnected networks on the eigenvalue centralities.}Here $%
\mathbf{E}_{ij}$ denotes the matrix with $1$ on $\left( i,j\right) $ and $%
\left( j,i\right) $ entries, $0$ on all the other entries.

\item[(ii)] Removing an existing link between $i$ and $j$: $\mathbf{C}=-%
\mathbf{E}_{ij}$.\footnote{%
For instance, \cite{Ballester2010} investigate the impact of removing a link
in a delinquent network.}

\item[(iii)] Removing all the links associated with a given player $i$: $%
\mathbf{C}=-\sum_{j:g_{ij}=1}\mathbf{E}_{ij}$.\footnote{\cite{Ballester2006}
study the impact of removing a single node from a criminal network. Also
see \cite{Zenou2014} for a recent survey of key players.}

\item[(iv)] Creating new links while removing existing links simultaneously.%
\footnote{%
For instance, \cite{Cai2017} show that business meetings, which help firms
build  social connections, have positive impacts on firm performance. 
\cite{Konig2014} study a model of network formation with new links added and
existing links removed dynamically. To study  efficient network design
with a fixed number of total links, \cite{Belhaj2016} analyze the effects of a 
\emph{link swap}, an operation  that cuts an existing link between $i$ and $%
j $ and  adds a new link between $k$ and $l$ (in our language, $\mathbf{C}%
=-\mathbf{E}_{ij}+\mathbf{E}_{kl}$ for a swap).}
\end{enumerate}
\end{ex}

Proposition \ref{prop-1} (i) describes the effect of interventions for each
player. In many applications, the designer may care about the aggregate
action or even its sign. Obviously, if links are created -- i.e., $\mathbf{C}%
\succeq \mathbf{0}$ -- the aggregate action unambiguously increases. Likewise,
when links are removed -- i.e., $\mathbf{C}\preceq \mathbf{0}$ -- the aggregate
action decreases. Suppose that new links are formed and meanwhile existing
links are removed in the intervention $\mathbf{C}$. Some players become more
active and others become less active, with the net effect on aggregate action
less clearcut to check. Proposition \ref{prop-1} (ii) provides a necessary
and sufficient condition. In the next Corollary, we present a sufficient
condition to guarantee that a structural intervention leads to higher
aggregate action. Such a condition is much simpler to check than that in
Proposition \ref{prop-1} (ii).\footnote{%
To use Proposition \ref{prop-1} (ii), we need to check the sign of $\mathbf{b%
}_{S}^{\prime }\delta \mathbf{C}_{SS}\left( \mathbf{I}-\delta \mathbf{M}%
_{SS}\left( \mathbf{G}\right) \mathbf{C}_{SS}\right) ^{-1}\mathbf{b}%
_{S}\left( \mathbf{G},\boldsymbol{\theta }\right) $.}

\begin{corr}
\label{cor-1} Assume $\boldsymbol{\theta}=\mathbf{1}$. In network $\left( N,%
\mathbf{G}\right) $, a structural intervention $\mathbf{C}$ that satisfies 
\begin{equation}  \label{eq:kkt}
\mathbf{b}^{\prime}\mathbf{C} \mathbf{b}= \mathbf{b}_{S}^{\prime}\mathbf{C}%
_{SS}\mathbf{b}_{S}\geq (>) 0
\end{equation}
always increases (strictly increases) aggregate equilibrium action, where $%
\mathbf{b}=\mathbf{b}\left( \mathbf{G},\mathbf{1},\delta\right)$.
\end{corr}

Corollary \ref{cor-1} is a direct consequence of the following inequality: 
\begin{equation}
\underbrace{b(\mathbf{G}+\mathbf{C})-b(\mathbf{G})}_{:=\Delta x^*} \geq
\delta \mathbf{b}^{\prime }\left( \mathbf{G}\right) \mathbf{C}\mathbf{b}(%
\mathbf{G}),
\end{equation}
which, under the condition $\boldsymbol{\theta}=\mathbf{1}$, provides a
lower bound on the change in aggregate action for any intervention $\mathbf{C%
}$ in $\Gamma (\mathbf{G},\mathbf{1})$. The above inequality employs a
convexity property of the equilibrium aggregation effort $b(\mathbf{G})$ as
a function of the network topology $\mathbf{G}$. Since the term on the
right-hand side can be viewed as the linear approximation ( and hence an
 underestimation due to convexity) of the change in
aggregate action. The condition stated in Corollary %
\ref{cor-1} is sufficient, but in general not necessary. An intervention
that does not satisfy the condition in Corollary \ref{cor-1} could still
improve aggregate action. Since $c_{ij}\in \{0, \pm 1\}$, we
can reformulate the expression in Corollary \ref{cor-1} as follows:
\begin{equation}
\mathbf{b}_{S}^{\prime }\left( \mathbf{G}\right) \mathbf{C}_{SS}\mathbf{b}%
_{S}\left( \mathbf{G}\right) =\sum_{i,j\in S} c_{ij}b_ib_j
=\left(\sum_{(i,j): c_{ij}=1} b_i b_j\right) - \left( \sum_{(i,j):
c_{ij}=-1} b_i b_j\right).
\end{equation}
If we define the product if tge Katz-Bonacich centralities of two nodes associated
with a link as the \emph{l-value} of that link, then Corollary \ref{cor-1}
states that if the sum of \emph{l-values} over new links in an intervention $%
\mathbf{C}$ exceeds the sum of \emph{l-values} over removed links in $%
\mathbf{C}$, the aggregate action must increase after this intervention. To
see some immediate implications of this Corollary, we present two simple
examples.

\begin{enumerate}
\item First, we consider a link reallocation in the form $\mathbf{C}=-%
\mathbf{E}_{ij}+\mathbf{E}_{kl}$ i.e., removing the link between $i$ and $j$%
and adding a new link between $k$ and $l$.\footnote{%
To make such an intervention $\mathbf{C}$ legitimate for network $\mathbf{G}$%
, we assume $g_{ij}=1$ and $g_{kl}=0$. Moreover, we assume at least three
elements of $\{i,j,k,l\}$ must be distinct.} Then Corollary \ref{cor-1}
implies that such a reallocation of links increases aggregate action if $b_i
b_j < b_k b_l$. In particular, it holds when $b_i<b_k$ and $b_j\leq b_l$.
Whenever the newly formed link contains nodes with higher Katz-Bonacich
centralities than the removed link, this type of link reallocation increases
aggregate action.

\item Second, we consider a link swap $\mathbf{\tilde C}=-\mathbf{E}_{ij}+%
\mathbf{E}_{il}$ (a specific reallocation with $i=k$), i.e., removing the
link between $i$ and $j$ and adding a new link between $i$ and $l$. Such
a swap, by Corollary \ref{cor-1}, increases aggregation action whenever $%
b_j<b_l$. Cutting an old link with a neighboring node $j$ of node $i$ with
lower Katz-Bonacich centrality and creating a new link from $i$ to
another unconnected node $l$ with higher Katz-Bonacich centrality makes the
whole group overall more active.
\end{enumerate}

In both examples, we identify simple ways to reallocate or swap links in an
existing network to improve aggregate action. This argument is complementary
to the critical Lemma (Lemma 1) in \citet{Belhaj2016}, which states that a
certain type of link swap or reallocation leads to higher aggregate welfare.%
\footnote{%
The planner's objective in \citet{Belhaj2016} is aggregate welfare, not
aggregate effort as in Corollary \ref{cor-1}. But the underlying driving
forces behind Lemma 1 in their paper are similar to ours. See our companion
paper, \cite{SZZ2021b}, for related discussions and further implications of
this Corollary on efficient network design.}

Another potential application of Corollary \ref{cor-1} is to provide local
optimality conditions for a constrained network optimization problem. Take a
set of networks $\mathcal{G}$. If $\mathbf{G}^{\ast }\in \mathcal{G}$ solves
the problem: $\max {x}^{\ast }(\mathbf{G^{\prime }})$ subject to $\mathbf{G}%
^{\prime }\in \mathcal{G}$. Then the optimality of $\mathbf{G}^{\ast }$
immediately implies that $\mathbf{b}^{\prime }\mathbf{C}\mathbf{b}\leq 0$ for
any $\mathbf{C}=\mathbf{G^{\prime }}-\mathbf{G}^{\ast }$ with $\mathbf{G}%
^{\prime }\in \mathcal{G}$, where $\mathbf{b}=\mathbf{b}\left( \mathbf{G}%
^{\ast }\right) $.\footnote{%
Since the network structure we consider in this paper is discrete (the
bilateral link is either zero or one), not continuous, the standard KKT
conditions for optimality do not directly apply. Focusing on the class of
weighted and directed networks, \cite{Li2019} employs KKT conditions to show
that optimal networks in his setting are generalized nested split graphs.}
For certain specifications of $\mathcal{G}$, the combinations of these local
necessary conditions are rich enough to infer useful structural properties
of the resulting optimal network.\footnote{%
In a companion paper, \cite{SZZ2021b}, on designing efficient networks
sequentially, we show that the optimal network $\mathbf{G}^{t}$ in each step 
$t$ must be contained in an important class of networks called quasi-complete
graphs.}

We mainly focus on the network model in \cite{Ballester2006}. As shown by 
\cite{Bramoulle2014}, our results regarding the effects of interventions on
networks carry over to a more general class of utility functions that induce
linear best responses.

Two main forms of interventions are studied in this section. Lemma \ref{lm-1}
focuses on general characteristic intervention. Proposition \ref{prop-1} and
Corollary \ref{cor-1} provide a unified approach to analyze the 
impacts of structural intervention on the equilibrium behavior at the individual and
aggregate level. The gist of Lemma \ref{lm-2} is to offer a new perspective
on identifying a structural intervention and an \emph{endogenously determined}
characteristic intervention. By restricting consideration to more specific types of
structural interventions in subsequent applications, we illustrate several
advantages of our theory of interventions in networks, compared with
existing approaches in the literature.

\section{The most wanted criminal(s) in delinquent networks}

\label{s-key group}

\subsection{The key group problem and the intercentrality index}

\label{sec:3.1}

Consider the following optimization problem: 
\begin{equation}
\underset{S\subseteq N,|S|\leq k}{\min }b\left( \mathbf{G}_{S^{C}S^{C}},%
\boldsymbol{\theta }_{S^{C}}\right) ,  \label{P-keygroup}
\end{equation}%
which is motivated by  application to a criminal network (see \cite%
{Ballester2006,Ballester2010} for detailed discussion): The government,
facing a group of criminals in a network $\mathbf{G}$, wants to identify a
subset of criminals $S$ of $N$ (known as \emph{the most wanted}) so that
the total action (criminal effort in this context) in the remaining network
is minimized after removing $S$ from the original network $\mathbf{G}$.%
\footnote{%
It is well established that criminality is a social action with strong peer
influences (see, for example, \cite{Sarnecki2001,Warr2002,Patacchini2012}).}
Note that before the intervention, the criminals play a game $\Gamma \left( 
\mathbf{G},\boldsymbol{\theta }\right) $ with total criminal activities $b\left( 
\mathbf{G},\boldsymbol{\theta }\right) $; after the removal of $S$ from $%
\mathbf{G}$, the remaining criminals $S^{C}$ play the game $\Gamma (\left( 
\mathbf{G}_{S^{C}S^{C}},\boldsymbol{\theta }_{S^{C}}\right) $, which leads
to the objective stated in program \eqref{P-keygroup}. To accommodate the
constraint from the government side (for instance, limited police resources),
the size of $S$ is bounded above by a positive integer $k$.

Problem \eqref{P-keygroup} is called the key player problem for $%
k=1$, and the key group problem in general for $k\geq 2$. \cite%
{Ballester2010} introduce the following definition.

\begin{defin}
\label{intercentrality} The intercentrality index of group $S$ in
network $\left( N,\mathbf{G}\right) $ is defined as%
\begin{equation*}
d_{S}\left( \mathbf{G},\boldsymbol{\theta }\right) \equiv b\left( \mathbf{G},%
\boldsymbol{\theta }\right) -b\left( \mathbf{G}_{S^{C}S^{C}},\boldsymbol{%
\theta }_{S^{C}}\right) \text{.}
\end{equation*}
\end{defin}

The intercentrality of group $S$, $d_{S}\left( \mathbf{G},\boldsymbol{\theta 
}\right) $, is the precise reduction of aggregate activity by removing
group $S$ and can be decomposed into two parts: a direct effect by the
removed players in $S$, $\sum_{l\in S}b_{l}\left( \mathbf{G},\boldsymbol{%
\theta }\right) $ and an  indirect effect due to the
decreasing  equilibrium actions of the remaining players $j\in S^{C}$, $%
\sum_{j\in S^{C}}\left\{ b_{j}\left( \mathbf{G},\boldsymbol{\theta }\right)
-b_{j}\left( \mathbf{G}_{S^{C}S^{C}},\boldsymbol{\theta }_{S^{C}}\right)
\right\} $. The next Lemma gives a simple expression for $d_{S}$. %

\begin{lem}
\label{key group} For any $S\subseteq N$, we have 
\begin{equation}
d_{S}\left( \mathbf{G},\boldsymbol{\theta }\right) =\mathbf{b}_{S}^{\prime
}\left( \mathbf{G}\right) \left( \mathbf{M}_{SS}\left( \mathbf{G}\right)
\right) ^{-1}\mathbf{b}_{S}\left( \mathbf{G},\boldsymbol{\theta }\right).
\label{KP index}
\end{equation}
\end{lem}

\begin{figure}[htp]
\begin{center}
\includegraphics[scale=0.5]{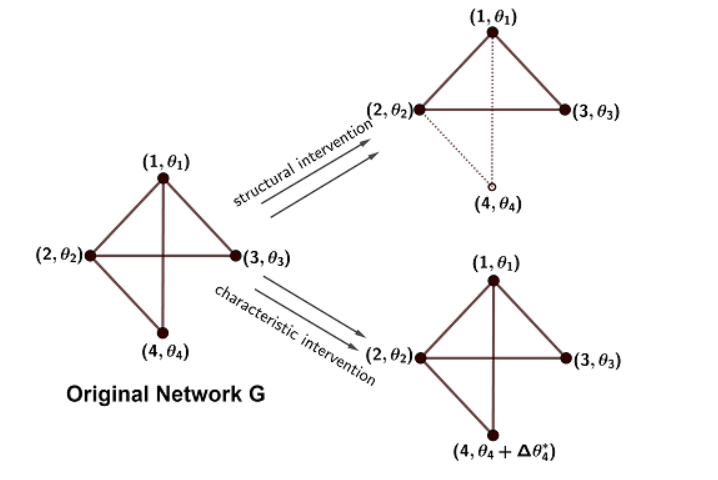}
\end{center}
\caption{Key player problem from the perspective of characteristic
intervention}
\label{fg-keygroup-equiv}
\end{figure}

For illustration, we consider Figure \ref{fg-keygroup-equiv}. 
 Imagine a thought experiment in which we change $\protect\theta%
_{4}$ by $\Delta \protect\theta _{4}^{\ast }=-\frac{b_{4}\left( \mathbf{G},%
\boldsymbol{\protect\theta }\right) }{m_{44}\left( \mathbf{G}\right) }$
while keeping other $\protect\theta _{i},i\neq 4$ the same. By Lemma \protect
\ref{lm-1}, after this characteristic intervention $\Delta \protect\theta %
_{4}^{\ast }$, each $i$'s equilibrium action is changed by $m_{i4}(\mathbf{G}%
)\Delta \protect\theta _{4}^{\ast }$ and the aggregate action is changed by 
$b_{4}(\mathbf{G})\Delta \protect\theta _{4}^{\ast }=-b_{4}(\mathbf{G},%
\mathbf{1})\frac{b_{4}\left( \mathbf{G},\boldsymbol{\protect\theta }\right) 
}{m_{44}\left( \mathbf{G}\right) }$. This critical value $\Delta \protect%
\theta _{4}^{\ast }$ is chosen so that player $4 $ will be exactly choosing
zero in equilibrium after this characteristic intervention: $b_{4}\left( 
\mathbf{G},\boldsymbol{\protect\theta }\right) +m_{44}(\mathbf{G})\Delta 
\protect\theta _{4}^{\ast }=0$. Given that player 4 is inactive in
equilibrium, the other three players effectively play a network game with
node $4$ removed. In other words, such a change of $\protect\theta _{4}$ by $%
\Delta \protect\theta _{4}^{\ast }$ exactly replicates the impacts of
removing $4$ from the network in terms of equilibrium choice. As a result, the total
impact of removing $4$ on the aggregate equilibrium activity is
given by $d_{4}(\mathbf{G},\boldsymbol{\protect%
\theta })=b_{4}(\mathbf{G},\mathbf{1})\frac{b_{4}\left( \mathbf{G},%
\boldsymbol{\protect\theta }\right) }{m_{44}\left( \mathbf{G}\right) }$.

The idea presented in Figure \ref{fg-keygroup-equiv} for a single node removal can
easily be extended to the setting with multiple nodes removed simultaneously.
The key observation is that the removal of group $S$ from the network has
the same effects as changing the characteristics of players in $S$ by 
\begin{equation*}
\Delta \boldsymbol{\theta }_{S}=-\left( \mathbf{M}_{SS}\left( \mathbf{G}
\right) \right) ^{-1}\mathbf{b}_{S}\left( \mathbf{G},\boldsymbol{\theta }
\right) \text{.}
\end{equation*}
According to equation \eqref{eq:aggregate partial}, this characteristic intervention leads to the
reduction of aggregate action by $-\mathbf{b}_{S}^{\prime }\left( \mathbf{G}
\right) \Delta \boldsymbol{\theta }_{S}=d_{S}(\mathbf{G},\boldsymbol{\theta }
)$ in equation \eqref{KP index}. 
Therefore, Lemma \ref{key group} follows immediately after showing the equivalence between 
a structural intervention (removal of a set of nodes) to a characteristic intervention
  (decreasing $\boldsymbol{\theta }_{S}$ by  $\Delta \boldsymbol{\theta }_{S}$). Consequently,
the key group program \eqref{P-keygroup} can be reformulated as 
\begin{equation}
\underset{S\subseteq N,|S|\leq k}{\max }d_{S}\left( \mathbf{G},\boldsymbol{
\theta }\right) =\mathbf{b}_{S}^{\prime }\left( \mathbf{G}\right) \left( 
\mathbf{M}_{SS}\left( \mathbf{G}\right) \right) ^{-1}\mathbf{b}_{S}\left( 
\mathbf{G},\boldsymbol{\theta }\right) .
\end{equation}

When $k=1$, taking $S=\left\{ i\right\} $, we obtain $d_{i}(\mathbf{G},
\boldsymbol{\theta })=\frac{b_{i}\left( \mathbf{G}\right) b_{i}\left( 
\mathbf{G},\boldsymbol{\theta }\right) }{m_{ii}\left( \mathbf{G}\right) }$
by equation \eqref{KP index}, which coincides with the
key player index in \cite{Ballester2006}. To the best of our knowledge, there is no analogous simple
expression for the key group index (or the intercentrality index) with $%
k\geq 2$, except for the definition. As a nontrivial generalization of
 the key player index, Lemma \ref{key group} uses the self-loops and centralities of the
removed players to construct the key group index. Thus, we can conveniently
identify the key group from the information in the matrix $\mathbf{M}\left( 
\mathbf{G}\right) $ without recomputing the new equilibrium after the
removal of nodes. Furthermore, the analytical simplicity of the expression
in Lemma \ref{key group} enables us to draw inference regarding   the key group.

\begin{pro}
\label{dominated groups} Assume $\boldsymbol{\theta} =\mathbf{1}$. Consider
two subsets, $S$ and $S^{\prime }$,

\begin{enumerate}
\item[(i)] if $S\subseteq (\subset)S^{\prime }$, then $d_{S}\left( \mathbf{G}%
,\boldsymbol{1}\right) \leq (<) d_{S^{\prime }}\left( \mathbf{G},\boldsymbol{%
1}\right) $;

\item[(ii)] if $\left\vert S\right\vert =\left\vert S^{\prime }\right\vert $%
, $\mathbf{b}_{S}\left( \mathbf{G}\right) \preceq \mathbf{b}_{S^{\prime
}}\left( \mathbf{G}\right) $, and $\mathbf{M}_{SS}\left( \mathbf{G}\right)
\succeq \mathbf{M}_{S^{\prime }S^{\prime }}\left( \mathbf{G}\right) $, then $%
d_{S}\left( \mathbf{G},\boldsymbol{1}\right) \leq d_{S^{\prime }}\left( 
\mathbf{G},\boldsymbol{1}\right) $.
\end{enumerate}
\end{pro}

Proposition \ref{dominated groups} (i) is rather intuitive: Removing a
larger group induces a more significant impact. In particular, to search for
the optimal $S^{\ast }$ in equation \eqref{P-keygroup}, it is without loss of 
generality to consider $S$ with $|S|=k$. Proposition \ref{dominated groups}
(ii) shows that when comparing groups of the same size, group $%
S^{\prime }$ with greater Katz-Bonacich centralities, $\mathbf{b}_{S^{\prime
}}$, and fewer  walks within any pair
of nodes in $S^{\prime }$ (measured by $\mathbf{M}_{S^{\prime }S^{\prime
}}\left( \mathbf{G}\right) $) has a larger intercentrality. These
monotonicity results are useful in reducing the possible choices of
candidates for the key group problem, as shown in Example \ref{ex2} below.%
\footnote{\cite{Ballester2010}, in their Example 2 on the key group
problem with $k=2$, illustrate similar observations as our Proposition \ref{dominated groups} (ii).}

\begin{ex}
\label{ex2} Consider a  regular network depicted in Figure %
\ref{fig-reg3}. We consider two cases: $k=1$ (the key player problem) and $%
k=2$ (the key group problem). Assume $\boldsymbol{\theta =1}$ and $%
\delta=0.2 $; then all nodes have the same unweighted Katz-Bonacich
centralities: $b_i(\mathbf{G},\mathbf{1})=b_j(\mathbf{G},\mathbf{1})$ for
any $i,j$.\footnote{%
This network is regular with degree $d=3$, so $\mathbf{G}^k\mathbf{1}= 3^k%
\mathbf{1},\forall k,$  and $b_i(\mathbf{G},%
\mathbf{1})=\frac{1}{1-d \delta}=2.5, \forall i$. This same network is also analyzed in 
\cite{calvo2004} and \cite{Zhou2015} under different contexts.}

\begin{figure}[tph]
\begin{center}
\includegraphics[scale=0.4]{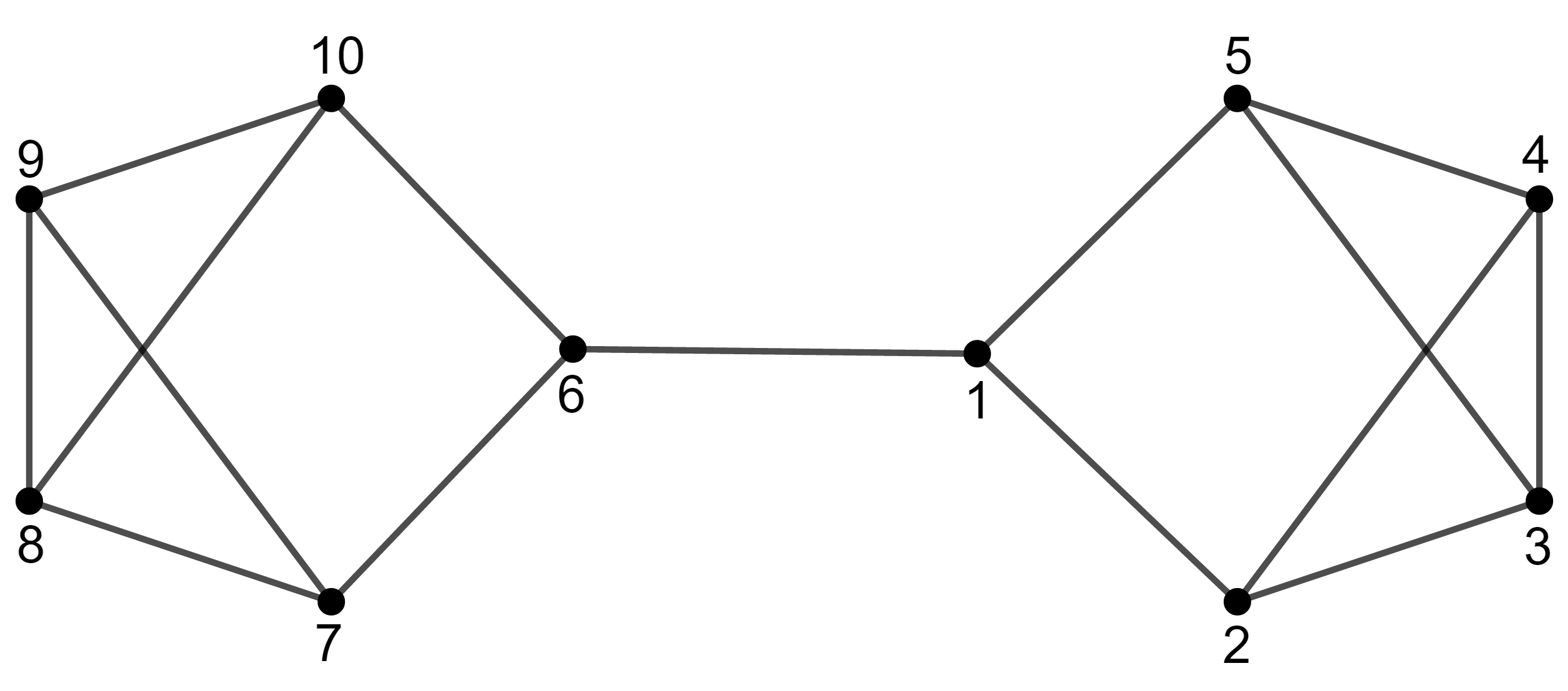}
\end{center}
\caption{A regular network with degree three}
\label{fig-reg3}
\end{figure}

\begin{enumerate}
\item[(i)] Assume $k=1$ so that we can only remove one node ($|S|=1$). For $%
S=\{i\}$, $d_i=\frac{b_i^2}{m_{ii}}$ by Lemma \ref{key group}. Since $b_i$
is the same for all $i$, so $d_l>d_j\Longleftrightarrow m_{ll}<m_{jj} $,
consistent with Proposition \ref{dominated groups} (ii). Table \ref{tb1}
summarizes $m_{ii}, d_i$ for each equivalent type of player.\footnote{%
For the key player problem ($k=1$), there are only three equivalent types
due to the symmetry of the network. For instance, players 1 and 6 are
equivalent. Similarly, $2,5,7$ and $10$ are mutually equivalent.} The key player
index is negatively related to self-loops. Therefore, player 1 (equivalently
player 6) is the key player.

\begin{table}[tph]
\caption{The key player}
\label{tb1}
\begin{center}
\begin{tabular}{ccc}
\hline\hline
$S$ & $m_{SS}\left( \mathbf{G}\right) $ & $d_{S}\left( \mathbf{G},%
\boldsymbol{1}\right) $ \\ \hline
\{1\} & 1.1688 & 5.3474* \\ 
\{2\} & 1.1981 & 5.2166 \\ 
\{3\} & 1.2162 & 5.1390 \\ \hline
\end{tabular}%
\end{center}
\end{table}
\begin{table}[tph]
\caption{The key group}
\label{tb2}
\begin{center}
\begin{tabular}{cc|cc}
\hline\hline
$S$ & $d_{S}\left( \mathbf{G},\boldsymbol{1}\right) $ & $S$ & $d_{S}\left( 
\mathbf{G},\boldsymbol{1}\right) $ \\ \hline
\{1,2\} & 8.4725 & \{2,3\} & 8.0331 \\ 
\{1,3\} & 9.3419 & \{2,5\} & 8.9529 \\ 
\{1,6\} & 8.7506 & \{2,7\} & 10.2938* \\ 
\{1,7\} & 10.0150 & \{2,8\} & 10.2863 \\ 
\{1,8\} & 10.2081 & \{3,4\} & 7.8174 \\ 
&  & \{3,8\} & 10.2431 \\ \hline
\end{tabular}%
\end{center}
\end{table}

\item[(ii)] Assume $k=2$ -- i.e., we can remove two nodes ($|S|=2$). Table \ref%
{tb2} shows the intercentralities of all equivalent types of groups with size $%
k=2$.\footnote{%
For this key group problem with $k=2$, there are exactly $11$ types up to
equivalence, as shown in Table \ref{tb2}. Note that node $2$ is equivalent to 
$7$ for the key player problem ($m_{22}=m_{77}$ and $b_2=b_7$), but $\{1,2\}$
and $\{1,7\}$ are not equivalent for the key group problem as $m_{12}\neq
m_{17}$.} By Proposition \ref{dominated groups} (ii), for $S=\{i,j\}$, $%
d_{S} $ is proportional to $\mathbf{1}^{\prime }\left( \mathbf{M}_{SS}
\right) ^{-1}\mathbf{1=}\frac{m_{ii}+m_{jj}-2m_{ij}}{m_{ii}m_{jj}-m_{ij}^{2}}
$ (note that $b_{i}=b_{j}$ for any $i,j$), and decreases in $m_{ii},m_{ij}$
and $m_{jj}$. Observe that $d_{\{2,7\}}$ is higher than $d_{\{2,5\}}$. Both
sets $\{2,7\}$ and $\{2,5\}$ share the same player 2; furthermore $m_{77}
=m_{55} =1.1981$, but $m_{27} =0.0162<m_{25} =0.1981$, implying $%
d_{\{2,7\}}>d_{\{2,5\}}$.\footnote{%
By the same token, we can show $d_{\{1,2\}}<d_{\{1,7\}}$, $%
d_{\{1,3\}}<d_{\{1,8\}}$, $d_{\{2,3\}}<d_{\{2,8\}}$.} In fact, as
demonstrated by Table \ref{tb2}, group $S^{\ast }=\left\{ 2,7\right\} $ is
the key group.
\end{enumerate}

For the key player problem with $k=1$, comparing the self-loop $m_{ii}$ is
sufficient to determine who is the most wanted player in the network in Figure %
\ref{fig-reg3}. The group version of intercentrality requires more detailed
information beyond self-loops. In particular, the number of walks between the
players in $S$, $m_{ij}$, also matters, as shown by the comparison between 
$\{2,7\}$ and $\{2,5\}$. Nevertheless, the monotonicity result in
Proposition \ref{dominated groups} (ii) enables us to rule out many
dominated groups for consideration.

Another interesting point is the comparison between $S^{\ast }=\left\{
2,7\right\} $ and $\hat S=\{1,6\}$. Both players 1 and 6 are key players with 
$k=1$ (see Table \ref{tb1}), but $\hat S=\{1,6\}$, the combination of two
key players, does not form the key group with $k=2$ as $d_{\{1,6\}}<d_{\{2,7%
\}}$.\footnote{%
Note that Proposition \ref{dominated groups} (ii) is not applicable here, as
the matrix $\mathbf{M}_{\hat S\hat S}$ does not dominate $\mathbf{M}_{S^*
S^*}$ entry by entry ($m_{22}=m_{77}>m_{11}=m_{66}$, but $m_{27}< m_{16}$).}
Simply collecting all of the key players together does not solve the key group
problem. In fact, in this example, the key group with $k=2$ does not include
any key player with $k=1$. These observations point to the computational
complexity of the key group problem, which is NP-hard (see Proposition 5 in 
\cite{Ballester2010} and the detailed discussion therein).
\end{ex}

\subsection{A view from walk counting}
 \label{sec:3.2}

Given the close relationship between equilibrium action in the game and
Katz-Bonacich centralities in the network, we offer an explanation of the
intercentrality index from the view of walk counting.

In network $\left( N,\mathbf{G}\right) $, $m_{ij}\left( \mathbf{G}\right) $
summarizes the total number of walks from $i$ to $j$ (with length discount $%
\delta $). In particular, for a non-empty set $S\subset N$, $m_{ij}\left( 
\mathbf{G}\right) $ counts the walks that pass one or more nodes in $S$, as
well as other walks that never hit any nodes in $S$. The former types of
walks with length discount exactly measure the importance of group $S$ in
the key group problem, since those walks do not contribute to the centrality
in the remaining network $\mathbf{G}_{S^{C}S^{C}}$. To distinguish these two
types of walks and facilitate the walk counting, we introduce the following
notation.

\begin{defin}
\label{def-W} Fixing a non-empty proper subset $S$ of $N$ in network $%
\left( N,\mathbf{G}\right) $, for any $i$, $j\in N$, we define $w_{ij}\left( 
\mathbf{G},S\right) $ as the total number of walks with length discount from $i$
to $j$ that do not pass any node in $S$, with the possible exception of the
starting node $i$ and the ending node $j$. Let $\mathbf{W}\left( \mathbf{G}%
,S\right) =\left( w_{ij}\left( \mathbf{G},S\right) \right) _{n\times n}$.
\end{defin}

Different from $m_{ij}\left( \mathbf{G}\right) $, $w_{ij}\left( \mathbf{G}%
,S\right) $ precludes the walks from $i$ to $j$ that cross group $S$. In
particular, $w_{ij}\left( \mathbf{G},\emptyset \right) =m_{ij}\left( \mathbf{%
G}\right) $. If $i$, $j\in S^{C}$, then $w_{ij}\left( \mathbf{G},S\right) $
counts the total number of walks from $i$ to $j$ that never pass group $S$%
; if $i\in S^{C}$ and $j\in S$, then $w_{ij}\left( \mathbf{G},S\right) $
counts the total number of walks from $i$ to $j$ that never pass group $S$
before stopping at node $j\in S$; if $i$, $j\in S$, then $w_{ij}\left( 
\mathbf{G},S\right) $ denotes the total number of walks from $i$ to $j$ that
never pass group $S$ except the starting and ending nodes $i$, $j$.

Since network $\mathbf{G}$ is undirected, matrix $\mathbf{W}\left( 
\mathbf{G},S\right) $ is necessarily symmetric: Any walk from $i$ to $j$
that bypasses group $S$ is also a walk from $j$ to $i$ that does not cross $%
S $, and vice versa. After suitable relabelling of nodes, $\mathbf{W}\left( 
\mathbf{G},S\right) $ can be represented as the following block matrix: 
\begin{equation*}
\mathbf{W}\left( \mathbf{G},S\right) =\left[ 
\begin{array}{cc}
\mathbf{W}_{S^{C}S^{C}}\left( \mathbf{G},S\right) & \mathbf{W}%
_{SS^{C}}\left( \mathbf{G},S\right) \\ 
\mathbf{W}_{S^{C}S}\left( \mathbf{G},S\right) & \mathbf{W}_{SS}\left( 
\mathbf{G},S\right)%
\end{array}%
\right] .
\end{equation*}%
The matrix $\mathbf{M}\left( \mathbf{G}\right) $ can be partitioned in the
same way. 
We establish the following result.

\begin{pro}
\label{prop-3} 
For any $S\subseteq N$, the following identities hold: 
\begin{eqnarray}
\mathbf{W}_{S^{C}S^{C}}\left(\mathbf{G},S\right) & = & \mathbf{M}_{S^{C}S^{C}}\left(\mathbf{G}\right)-\mathbf{M}_{S^{C}S}\left(\mathbf{G}\right)\left(\mathbf{M}_{SS}\left(\mathbf{G}\right)\right)^{-1}\mathbf{M}_{SS^{C}}\left(\mathbf{G}\right)\label{eq: Lemma 1 extension 1}\\
\mathbf{W}_{S^{C}S}\left(\mathbf{G},S\right) & = & \mathbf{M}_{S^{C}S}\left(\mathbf{G}\right)\left(\mathbf{M}_{SS}\left(\mathbf{G}\right)\right)^{-1}\label{eq:Lemma 1 extension 2}\\
\mathbf{W}_{SS}\left(\mathbf{G},S\right) & = & 2\mathbf{I}-\left(\mathbf{M}_{SS}\left(\mathbf{G}\right)\right)^{-1}\label{eq: Lemma 1 extension 3}
\end{eqnarray}
In particular, for any $A,B\subseteq N$ and $\ensuremath{A\cap B=\emptyset}$,
then 
\begin{equation}
\begin{aligned}\mathbf{W}_{AB}\left(\mathbf{G},A\cup B\right) & =\left(\mathbf{M}_{AA}\left(\mathbf{G}\right)\right)^{-1}\mathbf{M}_{AB}\left(\mathbf{G}\right)\left(\mathbf{W}_{BB}\left(\mathbf{G},A\right)\right)^{-1}\\
 & =\left(\mathbf{W}_{AA}\left(\mathbf{G},B\right)\right)^{-1}\mathbf{M}_{AB}\left(\mathbf{G}\right)\left(\mathbf{M}_{BB}\left(\mathbf{G}\right)\right)^{-1}
\end{aligned}\footnote{By equation \eqref{eq: Lemma 1 extension 1}, we further have 
 \[
\begin{aligned}\mathbf{W}_{AB}\left(\mathbf{G},A\cup B\right)= & \left(\mathbf{M}_{AA}\left(\mathbf{G}\right)\right)^{-1}\mathbf{M}_{AB}\left(\mathbf{G}\right)\left(\mathbf{M}_{BB}\left(\mathbf{G}\right)-\mathbf{M}_{BA}\left(\mathbf{G}\right)\left(\mathbf{M}_{AA}\left(\mathbf{G}\right)\right)^{-1}\mathbf{M}_{AB}\left(\mathbf{G}\right)\right)^{-1}\\
= & \left(\mathbf{M}_{AA}\left(\mathbf{G}\right)-\mathbf{M}_{AB}\left(\mathbf{G}\right)\left(\mathbf{M}_{BB}\left(\mathbf{G}\right)\right)^{-1}\mathbf{M}_{BA}\left(\mathbf{G}\right)\right)\mathbf{M}_{AB}\left(\mathbf{G}\right)\left(\mathbf{M}_{BB}\left(\mathbf{G}\right)\right)^{-1}
\end{aligned}
\]
 }
\label{eq:Lemma 1 extension 4}
\end{equation}
\end{pro}

Proposition \ref{prop-3} characterizes the impacts of removing group $S$ on
the total number of walks between each pair of nodes.\footnote{We provide a view from walk counting for the identities in Proposition \ref{prop-3} in Appendix \ref{app-bridge}.} Three points are
worth noting.

\begin{enumerate}
\item Equation \eqref{eq: Lemma 1 extension 1} uses centrality measures in
the original network to quantify all  of the walk changes in the remaining network when a
set of nodes is removed. Specifically, $\mathbf{M}_{S^{C}S^{C}}\left( 
\mathbf{G}\right) -\mathbf{W}_{S^{C}S^{C}}\left( \mathbf{G},S\right) =%
\mathbf{M}_{S^{C}S}\left( \mathbf{G}\right) \left( \mathbf{M}_{SS}\left( 
\mathbf{G}\right) \right) ^{-1}\mathbf{M}_{SS^{C}}\left( \mathbf{G}\right) $
summarizes the reduction in the total number of walks between each pair of
nodes in $S^{C}$. When $S=\left\{ i\right\} $, for any pair $\left(
j,k\right) \in S^{C}$, equation \eqref{eq: Lemma 1 extension 1} yields%
\begin{equation*}
{m_{jk}}({\mathbf{G}})-{w_{jk}}({\mathbf{G}},\{i\})=\frac{{{m_{ji}}\left( {%
\mathbf{G}}\right) {m_{ik}}\left( {\mathbf{G}}\right) }}{{{m_{ii}}\left( {%
\mathbf{G}}\right) }}\text{.}
\end{equation*}%
This equation is equivalent to Lemma 1 in \cite{Ballester2006}, which
characterizes the change in walks in the network after removing a single
node $i$ and leads to the intercentrality index. Equation \eqref{eq: Lemma 1
extension 1} extends  Lemma 1 in \cite{Ballester2006} to the case of
removing multiple nodes.

\item Equation \eqref{eq:Lemma 1 extension 2} indicates that the
intercentrality measure $d_{S}$ in equation \eqref{KP index} is precisely the
discounted number of walks that pass through group $S$. To fix this idea, we set $%
\mathbf{\theta =1}$. The intercentrality of group $S$ can be decomposed
according to whether the starting node of such a walk is in $S$ (type I
walks) or not (type II walks): 
\begin{equation}
d_{S}\left( \mathbf{G,1}\right) ={{\underbrace{b_{S}\left( \mathbf{G}\right) 
}_{\text{Term I}}+\underbrace{\mathbf{1}^{\prime }\mathbf{W}_{S^{C}S}(%
\mathbf{G},S)\mathbf{b}_{S}\left( \mathbf{G}\right) }_{\text{Term II}}}\text{%
.}} \footnote{%
Equation \eqref{ds-decomp} is consistent with  equation \eqref{KP index}
in Lemma \ref{key group} as, mathematically, 
\begin{eqnarray*}
d_{S}\left( \mathbf{G,1}\right) &=&\mathbf{b}_{S}^{\prime }\left( \mathbf{G}%
\right) \left( \mathbf{M}_{SS}\left( \mathbf{G}\right) \right) ^{-1}\mathbf{b%
}_{S}\left( \mathbf{G}\right) =\mathbf{1}^{\prime }\left[ \mathbf{M}%
_{S^{C}S}\left( \mathbf{G}\right) ,\mathbf{M}_{SS}\left( \mathbf{G}\right) %
\right] \left( \mathbf{M}_{SS}\left( \mathbf{G}\right) \right) ^{-1}\mathbf{b%
}_{S}\left( \mathbf{G}\right) \\
&=&\mathbf{1}^{\prime }\mathbf{M}_{SS}\left( \mathbf{G}\right) \left( 
\mathbf{M}_{SS}\left( \mathbf{G}\right) \right) ^{-1}\mathbf{b}_{S}\left( 
\mathbf{G}\right) +\mathbf{1}^{\prime }\mathbf{M}_{S^{C}S}\left( \mathbf{G}%
\right) \left( \mathbf{M}_{SS}\left( \mathbf{G}\right) \right) ^{-1}\mathbf{b%
}_{S}\left( \mathbf{G}\right) \\
&=&b_{S}\left( \mathbf{G}\right) +\mathbf{1}^{\prime }\mathbf{W}_{S^{C}S}(%
\mathbf{G},S)\mathbf{b}_{S}\left( \mathbf{G}\right) ,
\end{eqnarray*}%
where we use equation \eqref{eq:Lemma 1 extension 2} in the last equality.}
\label{ds-decomp}
\end{equation}
Term I is precisely the sum of walks with the starting node in $S$ --  i.e,
type I walks $b_{S}(\mathbf{G})=\sum_{i\in S}{b_{i}(\mathbf{G})}$. Term II
exactly captures the walks that start with a node in $S^{C}$ and pass
group $S$ at least once -- i.e., type II walks. Each walk of type II can be
decomposed as the concatenation of two walks: Consider an arbitrary walk
starting from node $i\in S^{C}$ and ending at $j$ (which may or may not in $%
S $), which passes the group $S$ at least once. Let $l\in S$ be the first
node  at which the walk meets group $S$. Then this walk can be \emph{uniquely}
decomposed as the concatenation of a walk from $i$ to $l$ and the other
walk from $l$ to $j$. The former category of walks never crosses group $S$
before ending, and therefore is summarized by $w_{il}\left( \mathbf{G}%
,S\right) $. The total number of the latter category of walks, with length
discount, is counted by $m_{lj}\left( \mathbf{G}\right) $. Consequently, the
number of type II walks from $i$ to $j$ is given by $\sum_{l\in
S}w_{il}\left( \mathbf{G},S\right) m_{lj}\left( \mathbf{G}\right) $. Summing
over indices $i\in S^{C}$ and $j\in N$, we obtain term II $\sum_{i\in
S^{C}}\sum_{j\in N}\sum_{l\in S}w_{il}(\mathbf{G},S)m_{lj}(\mathbf{G})=%
\mathbf{1}^{\prime }\mathbf{W}_{S^{C}S}(\mathbf{G},S)\mathbf{b}_{S}\left( 
\mathbf{G}\right) $. As a whole, the intercentrality of group $S$, $%
d_{S}\left( \mathbf{G,1}\right) $, is the discounted number of walks in
network $\left( N,\mathbf{G}\right) $ that pass group $S$ at least once.

\item Equation \eqref{eq: Lemma 1 extension 3} captures the aggregate walks
from $i$ to $j$ without passing any of them along the path. In particular,
let $S=\left\{ i,j\right\} $; then we have%
\begin{equation}
{w_{ij}}({\mathbf{G}},\{i,j\})=\frac{{{m_{ij}}({\mathbf{G}})}}{{{m_{ii}}({%
\mathbf{G}}){m_{jj}}({\mathbf{G}})-{{\left( {{m_{ij}}({\mathbf{G}})}\right) }%
^{2}}}}\text{.}\footnote{%
This equality is based on the following expression of inverse of block matrx%
\begin{equation*}
\left( \mathbf{M}_{SS}\left( \mathbf{G}\right) \right) ^{-1}=\left[ 
\begin{array}{cc}
m_{ii}\left( \mathbf{G}\right)  & m_{ij}\left( \mathbf{G}\right)  \\ 
m_{ji}\left( \mathbf{G}\right)  & m_{jj}\left( \mathbf{G}\right) 
\end{array}%
\right] ^{-1}=\frac{1}{m_{ii}\left( \mathbf{G}\right) -\frac{%
m_{ij}^{2}\left( \mathbf{G}\right) }{m_{jj}\left( \mathbf{G}\right) }}\left[ 
\begin{array}{cc}
1 & -\frac{m_{ij}\left( \mathbf{G}\right) }{m_{ii}\left( \mathbf{G}\right) }
\\ 
-\frac{m_{ij}\left( \mathbf{G}\right) }{m_{ii}\left( \mathbf{G}\right) } & 1%
\end{array}%
\right] \text{.}
\end{equation*}%
}  \label{eq:proposition2018}
\end{equation}%
Equation \eqref{eq:proposition2018} is consistent with Proposition 2 in \cite%
{Bramoulle2018} on targeting centralities, which characterizes the expected
number of times $i$'s request reaches $j$ if both nodes $i$ and $j$ are
excluded from favor retransmission.\footnote{%
The economic issue explored by \cite{Bramoulle2018} and the notation they
use differ slightly from ours. Here we have adapted their results
using our notation.} In fact, Proposition \ref{prop-3} generalizes \citeauthor%
{Bramoulle2018}'s targeting centrality in two dimensions. Specifically,
equation \eqref{eq: Lemma 1 extension 3} captures the expected times  $i$%
's request reaches $j$ when a group of individuals, rather than only $i$ and 
$j$ in \cite{Bramoulle2018}, are excluded from retransmission. Meanwhile,
equations \eqref{eq: Lemma 1 extension 1} and \eqref{eq:Lemma 1 extension 2}
capture the cases  in which either request originator $i$ or receiver $j$, or both,
are allowed to retransmit the request when a group of individuals cannot.
Finally, it is worth noting that equation \eqref{eq:Lemma 1 extension 4}
demonstrates a symmetric decomposition of the walks between two disjoint sets
of nodes in the network. This symmetric property is a group generalization
of \citeauthor{Bramoulle2018}'s observation (cf. footnote 7 in \cite{Bramoulle2018}%
).\footnote{\cite{Bramoulle2018} state that (in our notation):
\begin{quote}``For any $i$, $j$, $m_{ii}\left( \mathbf{G}\right) w_{jj}\left( \mathbf{G}%
,\left\{ i\right\} \right) =$ $m_{jj}\left( \mathbf{G}\right) w_{ii}\left( 
\mathbf{G},\left\{ j\right\} \right) $. To our knowledge, this provides a
noval result in matrix analysis.''\end{quote}
 The symmetric property is consistent with \eqref{eq:Lemma 1 extension 4} after canceling 
 the common term $m_{ij}\left( \mathbf{G}\right) $ on both sides.}
\end{enumerate}

\section{The key bridge connecting isolated networks}

\label{s-Bridge}

\subsection{The bridge index and the key bridge}

\label{sec:4.1}

Consider two isolated networks, $\left( N_{1},\mathbf{N}^{1}\right) $ and $%
\left( N_{2},\mathbf{N}^{2}\right) $, where $N_{i}$ denotes the set of
players and $\mathbf{N}^{i}$ denotes the corresponding adjacency matrix for $%
i\in \left\{ 1,2\right\} $. Define $N=N^{1}\cup N^{2}$ and the adjacency
matrix $\mathbf{G}=%
\begin{bmatrix}
\mathbf{N}^{1} & \mathbf{0} \\ 
\mathbf{0} & \mathbf{N}^{2}%
\end{bmatrix}%
$. Note that $m_{ij}(\mathbf{G})=m_{ji}(\mathbf{G})=0$ for $i\in N_{1},j\in
N_{2}$. To simplify the notation, we set $\theta _{k}=1$ for any 
$k\in N$ throughout this section. The planner's problem is
to maximize aggregate equilibrium effort by adding a new link between some
nodes $i\in N_{1}$ and $j\in N_{2}$. Mathematically, the planner solves%
\begin{equation}
\max_{(i,j)\in N_{1}\times N_{2}}b\left( \mathbf{G+E}_{ij}\right) \text{.}
\label{KBP problem}
\end{equation}%
The pair of nodes $(i^{\ast },j^{\ast })$ that solves the above problem is
called the \textit{key bridge pair}. We call node $i^{\ast }$ (node $j^{\ast
}$)  the key bridge player in network $\mathbf{N}^{1}(\mathbf{N}^{2})$.

The key bridge problem naturally arises in many economic settings. For
example, in the integration of new immigrants into a new country, the
communication between cultural leaders serves as a bond that connects two
initially isolated communities (see \cite{VERDIER2015,VERDIER2018}). For
another example, a firm can be viewed as a network among workers with
synergies, since a worker's productivity is influenced by his peers through
knowledge sharing and skill complementarity. Building  interfirm social
connections creates further economic value. For instance, \cite{Cai2017}
document the effects of interfirm meetings  between young Chinese firms on
their business performance.\footnote{%
In a large-scale experimental study of network formation, \citet{choi2020}
highlight the role of connectors and influencers.}

\begin{defin}
\label{Bridge index} For any pair $(i,j)\in N_1\times N_2$, define the
bridge index $L_{ij}\left( \mathbf{N}^{1},\mathbf{N}^{2}\right) $ as 
\begin{equation}
L_{ij}\left( \mathbf{N}^{1},\mathbf{N}^{2}\right) \equiv \frac{\delta
m_{jj}\left( \mathbf{N}^{2}\right) b_{i}^{2}\left( \mathbf{N}^{1}\right)
+\delta m_{ii}\left( \mathbf{N}^{1}\right) b_{j}^{2}\left( \mathbf{N}%
^{2}\right) +2b_{j}\left( \mathbf{N}^{2}\right) b_{i}\left( \mathbf{N}%
^{1}\right) }{1-\delta ^{2}m_{jj}\left( \mathbf{N}^{2}\right) m_{ii}\left( 
\mathbf{N}^{1}\right) }.  \label{Bridge Index}
\end{equation}
\end{defin}

\begin{pro}
\label{pro-bridge} The key bridge pair $(i^*,j^*)$ must maximize the bridge
index, i.e., 
\begin{equation*}
(i^*,j^*)\in \arg\max_{(i,j)\in N_1\times N_2} L_{ij}\left( \mathbf{N}^{1},%
\mathbf{N}^{2}\right) \text{.}
\end{equation*}
\end{pro}

This Proposition fully solves the key bridge problem using the bridge index.
It follows that when $\mathbf{G}$ is the union of two isolated networks,
\begin{equation}  \label{plus-ij}
b\left( \mathbf{G+E}_{ij}\right)-b\left( \mathbf{G}\right) =\delta
L_{ij}\left( \mathbf{N}^{1},\mathbf{N}^{2}\right), ~~~ \forall i\in N_1,
j\in N_2.
\end{equation}
Thus, this bridge index $L_{ij}$ summarizes all of the walks passing bridge 
$(i,j)$ at least once. Depending on the starting node, the end node, and how
many times such a new walk intersects with the bridge, we sort these
additional walks into different categories and can provide a view from walk
counting for each category, as shown in the bridge index (see  Appendix %
\ref{app-bridge} for details).

To obtain further insights on exactly who is the key bridge player using 
primitive information, we present the following Corollary. Let $%
e_{i}=\sum_{j\in N}g_{ij}$ denote the degree of player $i$.

\begin{corr}
\label{bridge cor} The following properties of the bridge index $%
L_{ij}\left( \mathbf{N}^{1},\mathbf{N}^{2}\right) $ hold:

\begin{enumerate}
\item[(i)] For two nodes $i$ and $i^{\prime }$ in $N_{1}$ with $b_{i}\left( 
\mathbf{N}^{1}\right) \geq b_{i^{\prime }}\left( \mathbf{N}^{1}\right) $ and 
$m_{ii}\left( \mathbf{N}^{1}\right) \geq m_{i^{\prime }i^{\prime }}\left( 
\mathbf{N}^{1}\right) $, we have $L_{ij}\left( \mathbf{N}^{1},\mathbf{N}%
^{2}\right) \geq L_{i^{\prime }j}\left( \mathbf{N}^{1},\mathbf{N}^{2}\right) 
$ for any $j\in N_{2}$.

\item[(ii)] For nodes $i$, $i^{\prime }$ in $N_{1}$ and $j$, $j^{\prime }$
in $N_{2}$ such that $b_{i}\left( \mathbf{N}^{1}\right) =b_{i^{\prime
}}\left( \mathbf{N}^{1}\right) $, $m_{jj}\left( \mathbf{N}^{2}\right)
=m_{j^{\prime }j^{\prime }}\left( \mathbf{N}^{2}\right) $, $m_{ii}\left( 
\mathbf{N}^{1}\right) \geq m_{i^{\prime }i^{\prime }}\left( \mathbf{N}%
^{1}\right) $ and $b_{j}\left( \mathbf{N}^{2}\right) \geq b_{j^{\prime
}}\left( \mathbf{N}^{2}\right) $, we have 
\begin{equation*}
L_{ij}\left( \mathbf{N}^{1},\mathbf{N}^{2}\right) -L_{ij^{\prime }}\left( 
\mathbf{N}^{1},\mathbf{N}^{2}\right) \geq L_{i^{\prime }j}\left( \mathbf{N}%
^{1},\mathbf{N}^{2}\right) -L_{i^{\prime }j^{\prime }}\left( \mathbf{N}^{1},%
\mathbf{N}^{2}\right) \text{.}
\end{equation*}

\item[(iii)] For two nodes $i$, $i^{\prime }$ in $N_{1}$ with $%
e_{i}>e_{i^{\prime }}$, there exists $\bar{\delta}>0$ such that for any $%
0<\delta <\bar{\delta}$, $L_{ij}\left( \mathbf{N}^{1},\mathbf{N}^{2}\right)
> L_{i^{\prime }j}\left( \mathbf{N}^{1},\mathbf{N}^{2}\right) $ for any $%
j\in N_{2}$.
\end{enumerate}
\end{corr}

Corollary \ref{bridge cor} (i) implies that to find key bridge player $%
i^{\ast }$ in the first network,$N_{1}$, it suffices to focus on the set of
nodes in $N_{1}$ that lie on the Pareto frontier of Katz-Bonacich
centrality $b_{i}(\mathbf{N}^{1})$ and self-loops $m_{ii}(\mathbf{N}^{1})$.
Therefore, if the most active player (the one with highest $b_{i}(\mathbf{N}%
^{1})$) also happens to be the one with the highest $m_{ii}(\mathbf{N}^{1})$,
it must be the key bridge player. The reason is that $L_{ij}\left( \mathbf{N}%
^{1},\mathbf{N}^{2}\right) $ increases with $b_{i}(\mathbf{N}^{1})$ and $%
m_{ii}(\mathbf{N}^{1})$, fixing $j$. Meanwhile, the intercentrality index, $%
\frac{b_{i}^{2}(\mathbf{N}^{1})}{m_{ii}(\mathbf{N}^{1})}$, increases in $%
b_{i}(\mathbf{N}^{1})$ but decreases in $m_{ii}(\mathbf{N}^{1})$. Therefore,
the key player may differ from the key bridge player.

Moreover, when the player with the largest $b_{i}(\mathbf{N}^{1})$ differs from
the one with the largest $m_{ii}(\mathbf{N}^{1})$ in $N_{1}$, the selection of
the key bridge player in $N_{1}$ crucially depends on who is chosen as the
bridge player $j$ in the other network $\mathbf{N}^{2}$. In other words, the
selection of the key bridge pair is not independent. As suggested by
Corollary \ref{bridge cor} (ii), the role of node $i$'s self-loops in
determining the bridge index $L_{ij}\left( \mathbf{N}^{1},\mathbf{N}%
^{2}\right) $ becomes more significant if a node $j$ with larger
Katz-Bonacich centrality is selected. Correspondingly, $i$'s Katz-Bonacich
centrality plays a more important role than self-loops when a node $j$ with
larger self-loops is selected. Consider a scenario in which these two
networks are highly unbalanced, in the sense that the largest Bonacich
centrality in network $\mathbf{N}^{2}$ is much larger than the one in
network ${\mathbf{N}^{1}}$.\footnote{%
This could be the case when network $\mathbf{N}^{2}$ involves much more
network members, who have denser connections.} Then the key
bridge pair may consist of the central node in $\mathbf{N}^{2}$ and the node
with the largest self-loop in network $\mathbf{N}^{1}$ (even if this node may
not be the most central one). This is because the self-loop of the
player in $\mathbf{N}^{1}$ contributes more than Katz-Bonacich centrality to
the bridge index $L_{ij}\left( \mathbf{N}^{1},\mathbf{N}^{2}\right) $ when the
Katz-Bonacich centrality of the bridge player in $\mathbf{N}^{2}$ is
significantly large. Furthermore, when $\delta $ is below a threshold, the
degree centrality plays the dominant role in the bridge index by item (iii).
We illustrate these observations using the following example.

\begin{ex}
\label{ex-3} Consider the two isolated networks depicted in Figure \ref{fg: key
bridge}.

\begin{figure}[htp]
\centering
\par
\includegraphics[scale=0.49]{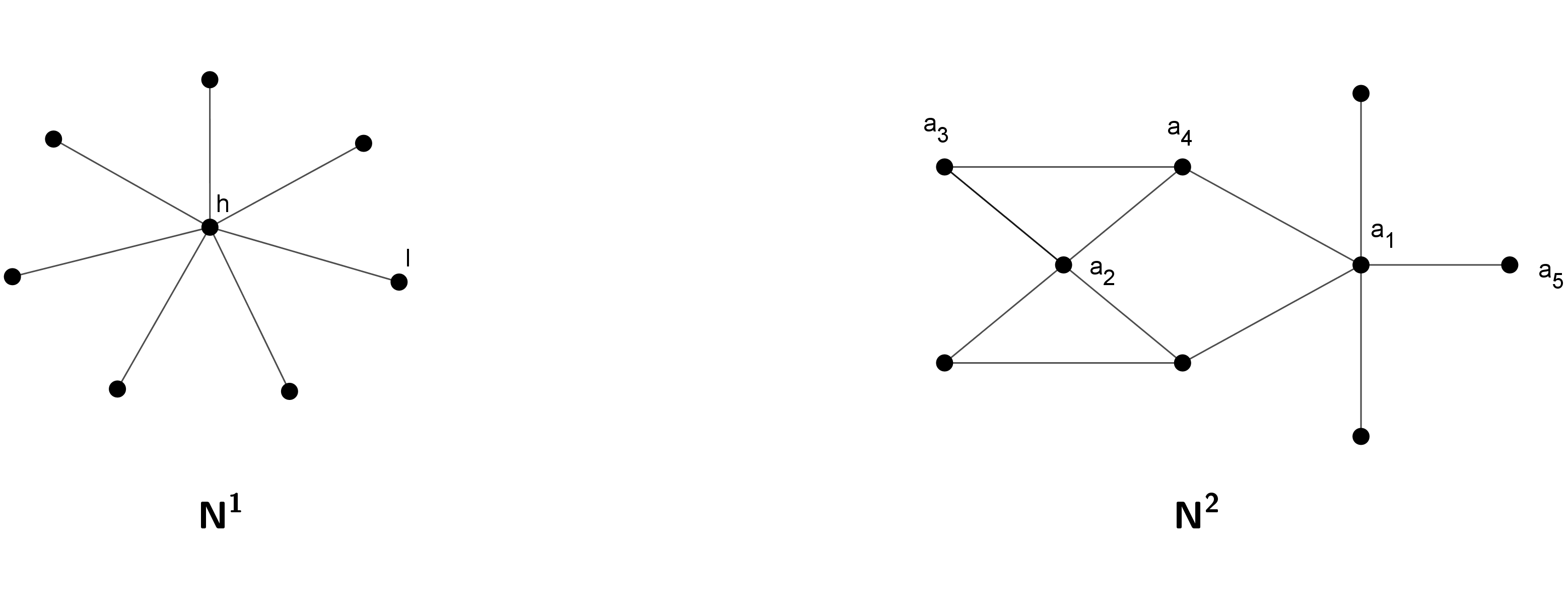}
\caption{Connecting $\mathbf{N}^{1}$ and $\mathbf{N}^{2}$ by adding a bridge}
\label{fg: key bridge}
\end{figure}

\begin{table}[htp]
\begin{minipage}{0.49\textwidth}
		\caption{Measures $\delta =0.25$}
		\begin{center}
\begin{tabular}{ccc}
	\hline\hline
	Players & $m_{ii}$ & $b_{i}$ \\ \hline
	$a_{1}$ & 1.5686 & \textbf{4.7059} \\ 
	$a_{2}$ & \textbf{1.5980} & 4.6765 \\ 
	$a_{3}$ & 1.2686 & 3.2059 \\ 
	$a_{4}$ & 1.4255 & 4.1471 \\ 
	$a_{5}$ & 1.0980 & 2.1765 \\ \hline
	$h$ & \textbf{1.7778} & \textbf{4.8889} \\ 
	$l$ & 1.1111 & 2.2222 \\ \hline
	\label{tab3}
\end{tabular}
	\end{center}
\end{minipage}%
\begin{minipage}{0.5\textwidth}
		\caption{Bridges with $\delta =0.25$}
		\begin{center}
\begin{tabular}{lc}
	 \hline\hline
	\multicolumn{1}{c}{bridge $i$-$j$} & $L_{ij}\left( \mathbf{G}\right) $ \\ \hline
	\multicolumn{1}{c}{$h$-$a_{1}$} & 78.9970 \\ 
	\multicolumn{1}{c}{$h$-$a_{2}$} & \textbf{79.0258} \\ \hline
	\multicolumn{1}{c}{} &  \\ 
	\multicolumn{1}{c}{} &  \\ 
	\multicolumn{1}{c}{} &  \\ 
	\multicolumn{1}{c}{} &  \\ 
	\multicolumn{1}{c}{} & 
	\label{tab4}
\end{tabular}
\end{center}
\end{minipage}
\end{table}

Table \ref{tab3} gives the Katz-Bonacich centrality $b_i$ and self-loops $%
m_{ii}$ measures for $\delta =0.25$. In the first network $\left( N_{1},%
\mathbf{N}^{1}\right) $, the hub player $h$ is more important than any of
the peripheral nodes both in terms of centrality and self-loop measures. In
the second network $\left( N_{2},\mathbf{N}^{2}\right) $, $a_1$ dominates $%
a_2$ in terms of Katz-Bonacich centrality $b_i$, while $a_2$ dominates $a_1$
in terms of self-loops $m_{ii}$. All other nodes in $N_2$ are dominated
by $a_1$ and $a_2$ in both $m_{ii}$ and $b_i$. By Corollary \ref{bridge cor}
(i), the key bridge pair is either $(h, a_1)$ or $(h, a_2)$. Table \ref{tab4}
demonstrates that  bridge index $L_{ha_{2}}\left( \mathbf{N}^{1},\mathbf{N%
}^{2}\right)> L_{ha_{1}}\left( \mathbf{N}^{1},\mathbf{N}^{2}\right) $, and thus $%
a_2$ is the key bridge player in $\left( N_{2},\mathbf{N}^{2}\right) $, yet $%
a_2$ is neither the most active player (in terms of $b_i$) nor the key
player (in terms of the intercentrality $b_i^2/m_{ii}$) in $\left( N_{2},%
\mathbf{N}^{2}\right)$.

Next, we consider $\delta=0.23$ (see Tables \ref{tab5} and \ref{tab6}). By
the same logic, it suffices to consider connecting the hub $h$ in the first
network to either $a_1$ or $a_2$ in the second network. However, for $%
\delta=0.23$, the $b_i$ plays a more prominent role in the bridge index than 
$m_{ii}$, and indeed $a_1$ is now the key bridge player. This observation is
consistent with Corollary \ref{bridge cor} (iii): The key bridge player is
the player with the highest degree when $\delta $ is relatively small (the
degree of $a_{1}$ is larger than that of $a_{2}$).

Keeping $\delta =0.23$. Suppose we increase the peripheral
nodes in $\left( N_{1},\mathbf{N}^{1}\right) $ from $7$ to $17$. The
Katz-Bonacich centrality of hub player $b_{h} =48.76$ is significantly
larger than that of players in $N_{2}$. Thus, the self-loops $m_{jj}$ of the bridge player $j$
in $N_{2}$ are more pronounced compared with his Katz-Bonacich centrality $%
b_{j}$ in shaping the relative values of $L_{hj}$. As a result, $(h,a_{2})$
is the key bridge (indeed, $L_{ha_{2}}\left( \mathbf{N}^{1},\mathbf{N}%
^{2}\right) =4744>L_{ha_{1}}\left( \mathbf{N}^{1},\mathbf{N}^{2}\right)
=4680 $).

\begin{table}[tbp]
\begin{minipage}{0.5\textwidth}
		\caption{Measures with $\delta =0.23$}
		\begin{center}	
\begin{tabular}{ccc}
	\hline\hline
	Players & $m_{ii}$ & $b_{i}$ \\ \hline
	$a_{1}$ & 1.4213 & \textbf{3.8423} \\ 
	$a_{2}$ & \textbf{1.4300} & 3.7545 \\ 
	$a_{3}$ & 1.1969 & 2.6348 \\ 
	$a_{4}$ & 1.3063 & 3.3533 \\ 
	$a_{5}$ & 1.0752 & 1.8837 \\ \hline
	$h$ & \textbf{1.5881} & \textbf{4.1448} \\ 
	$l$ & 1.0840 & 1.9533 \\ \hline
	\label{tab5}
\end{tabular}
	\end{center}
\end{minipage}%
\begin{minipage}{0.5\textwidth}
		\caption{Bridges with $\delta =0.23$}
		\begin{center}
\begin{tabular}{lc}
	 \hline\hline
	\multicolumn{1}{c}{Bridge $i$-$j$} & $L_{ij}\left( \mathbf{G}\right) $ \\ \hline
	\multicolumn{1}{c}{$h$-$a_{1}$} & \textbf{48.6711} \\ 
	\multicolumn{1}{c}{$h$-$a_{2}$} & 47.6461 \\ \hline
	\multicolumn{1}{c}{} &  \\ 
	\multicolumn{1}{c}{} &  \\ 
	\multicolumn{1}{c}{} &  \\ 
	\multicolumn{1}{c}{} &  \\ 
	\multicolumn{1}{c}{} &
	\label{tab6}
\end{tabular}
\end{center}
\end{minipage}
\end{table}
\end{ex}


\subsection{The value of an existing link and the value of a potential link}

\label{sec:4.2}

Instead of considering two isolated networks, in this subsection we consider
a general network and examine the effects of link creation and deletion.

\begin{lem}
\label{lm-keylink} For any network $\mathbf{G}$,

\begin{itemize}
\item[(i)] Suppose $g_{ij}=0$, then 
\begin{equation}  \label{L-ij}
b(\mathbf{G}+\mathbf{E}_{ij})-b(\mathbf{G})=\delta {L}_{ij}\left( \mathbf{G}%
\right)
\end{equation}
where 
\begin{equation*}
{L}_{ij}\left( \mathbf{G}\right) =\frac{\delta m_{ii}\left( \mathbf{G}%
\right) b_{j}^{2}\left( \mathbf{G}\right) +\delta m_{jj}\left( \mathbf{G}%
\right) b_{i}^{2}\left( \mathbf{G}\right) +2\left( 1-\delta m_{ij}\left( 
\mathbf{G}\right) \right) b_{i}\left( \mathbf{G}\right) b_{j}\left( \mathbf{G%
}\right) }{\left( 1-\delta m_{ij}\left( \mathbf{G}\right) \right)
^{2}-\delta ^{2}m_{ii}\left( \mathbf{G}\right) m_{jj}\left( \mathbf{G}%
\right) }\text{.}
\end{equation*}

\item[(ii)] Suppose $g_{ij}=1$, then 
\begin{equation}  \label{l-ij}
b(\mathbf{G}-\mathbf{E}_{ij})-b(\mathbf{G})=-\delta l_{ij}\left( \mathbf{G}%
\right)
\end{equation}
where 
\begin{equation*}
l_{ij}\left( \mathbf{G}\right) =\frac{2\left( 1+\delta m_{ij}\left( \mathbf{G%
}\right) \right) b_{i}\left( \mathbf{G}\right) b_{j}\left( \mathbf{G}\right)
-\left( \delta m_{ii}\left( \mathbf{G}\right) b_{j}^{2}\left( \mathbf{G}%
\right) +\delta m_{jj}\left( \mathbf{G}\right) b_{i}^{2}\left( \mathbf{G}%
\right) \right) }{\left( 1+\delta m_{ij}\left( \mathbf{G}\right) \right)
^{2}-\delta ^{2}m_{ii}\left( \mathbf{G}\right) m_{jj}\left( \mathbf{G}%
\right) }\text{.}
\end{equation*}
\end{itemize}
\end{lem}

The index $L_{ij}$ measures the value of a potential new link $(i,j)$ in the
network $\mathbf{G}$, while $l_{ij}$ measures the value of an existing link $%
(i,j)$ in $\mathbf{G}$. Therefore, the index $L_{ij}$ is useful for
determining the optimal location for a new link (for instance, the key
bridge problem). Meanwhile, $l_{ij}$ measures the contribution of an
existing link $\left( i,j\right) $ to the total Katz-Bonacich centralities.
For instance, \cite{Ballester2010} derive the same measure using a different
method (see their Lemma 2), and use it to study the most important existing link
(the key link). Both indices $L_{ij}$ and $l_{ij}$ share some similar
properties  with the bridge index in Corollary \ref{bridge cor}, and hence we omit
the details.

Both results in Lemma \ref{lm-keylink} follow directly from Proposition \ref%
{prop-1}: We set $\mathbf{C}=\mathbf{E}_{ij}$ in case (i) and set $\mathbf{C}%
=-\mathbf{E}_{ij}$ in case (ii). Note that in both cases, $m_{ij}$ and $b_i$
in the expressions of $L_{ij}$ and $l_{ij}$ are evaluated at the existing
network $\mathbf{G}$. Since removing the newly added link $(i,j)$ in $%
\mathbf{G}+\mathbf{E}_{ij}$ results in the original network $\mathbf{G}$,
Lemma \ref{lm-keylink} reveals the following relationship between two
indices: 
\begin{equation}
l_{ij}(\mathbf{G}+\mathbf{E}_{ij})=L_{ij}(\mathbf{G}).
\end{equation}
This identity enables us to express $L_{ij}(\mathbf{G})$ using the
centrality measures of the new network $\mathbf{G}+\mathbf{E}_{ij}$. Lemma %
\ref{lm-keylink} (i) generalizes the bridge index in equation \eqref{Bridge Index}.
Indeed, when $\mathbf{G}$ is the union of two isolated networks $\mathbf{N}
^1 $ and $\mathbf{N}^2$, we have $m_{ij}=0$ for $i\in \mathbf{N}^1,j\in
\mathbf{N}^2$, and the index $L_{ij}$ in equation \eqref{L-ij} reduces to $L_{ij}(
\mathbf{N}^1,\mathbf{N}^2)$ in equation \eqref{Bridge Index}.

Unlike the bridge index that does not depend on $m_{ij}$, the general link
 index $L_{ij}$  increases with $m_{ij}$ (note that in a general network, 
 $m_{ij}$ can be positive even when $i$ and $j$ are not directly connected).  
 So the designer may prefer connecting  nodes that have already been well 
 connected in the original network to adding bridges connecting disjointed
  groups. The next example illustrates when it is desirable to add 
  intragroup link(s) vs. intergroup link(s).

\begin{ex}
    \label{ex-4}
    Consider an original network $\mathbf{G}$ composed of two disjointed cycles, each of size four.
      We search for  the optimal way to add one or two links, in which
       the links can be formed  between or within two circles. 
    
    \begin{itemize}
    \item[(a)]
        Consider adding one link.  By symmetry, it suffices to compare   $\hat{\mathbf{G}}_1$ and $\hat{\mathbf{G}}_2$ in Figure \ref{fg:single_link_add}.  Given  $m_{2,3}>m_{2,5}=0$ and all other measures being equal, adding the intragroup link $(2,3)$ strictly dominates  adding the  inter-group link $(2,5)$; i.e., $\hat{\mathbf{G}}_2$ dominates $\hat{\mathbf{G}}_1$ (see Table  \ref{tab:centrality_comparison}).
     
    \begin{figure}[htp]
    \centering
    \par
    \includegraphics[scale=0.49]{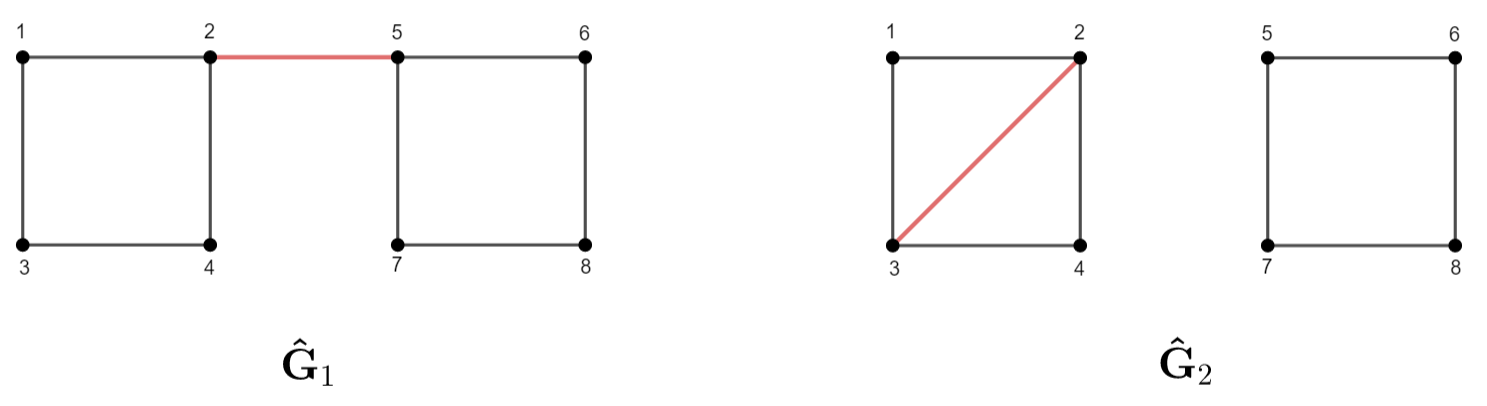}
    \caption{Adding a single link between two separated networks}
    \label{fg:single_link_add}
    \end{figure}

  \item[(b)]
   What if we can add one additional link?   It is easy to see that the optimal network is one
   of the following three networks: $\bar{\mathbf{G}}_1$, $\bar{\mathbf{G}}_2$, and $\bar{\mathbf{G}}_3$ (see Figure \ref{fg:two_links_add}).\footnote{All other ways of forming two links are dominated.  Denote $((i,j),(k,l))$ as an intervention of adding two links. For instance, $((2,5),(4,7))$ is strictly dominated by $((2,5),(2,7))$, since $b_2>b_4$, $m_{2,2}>m_{4,4}$ and $m_{2,7}>m_{4,7}$ once a bridge $(2,5)$ is added. $((2,5),(2,7))$ strictly dominates $((2,5),(2,8))$, since $m_{2,7}>m_{2,8}$ once a bridge $(2,5)$ is added. $((2,3),(6,7))$ is strictly dominated by $((1,4),(2,3))$, since $b_1>b_6$, $m_{1,1}>m_{6,6}$ and $m_{1,4}>m_{6,7}$ once $(2,3)$ is added. $((2,5),(6,7))$ is strictly dominated by $((5,8),(6,7))$, since $b_8>b_2$, $m_{8,8}>m_{2,2}$ and $m_{5,8}>m_{2,5}$ once $(6,7)$ is added.}  
   Of these three, Table \ref{tab:centrality_comparison} shows that  $\bar{\mathbf{G}}_3$ is the optimal.   In other words,   connecting two intergroup bridges $((2,5),(2,7))$ strictly dominates building  two intragroup links $((1,4),(2,3))$, even though building  one intragroup link is myopically optimal, as in part (a).\footnote{Starting with $\hat{\mathbf{G}}_2$, the optimal network with one extra link by part (a).  Conditioning on  adding $(2,3)$ as the first link,  $(1,4)$ strictly dominates $(2,5)$ as the second link, since  $\bar{\mathbf{G}}_2$ has higher aggregate Katz-Bonacich centralities  than $\bar{\mathbf{G}}_1$ in terms of  (see Table \ref{tab:centrality_comparison}).  However, neither $\bar{\mathbf{G}}_1$ nor $\bar{\mathbf{G}}_2$ is optimal. Nevertheless, starting with the  dominated network, $\hat{\mathbf{G}}_1$ in part (a), we can reach the optimal network  $\bar{\mathbf{G}}_3$  by adding the link $(2,7)$.}

    \begin{figure}[htp]
    \centering
    \par
    \includegraphics[scale=0.45]{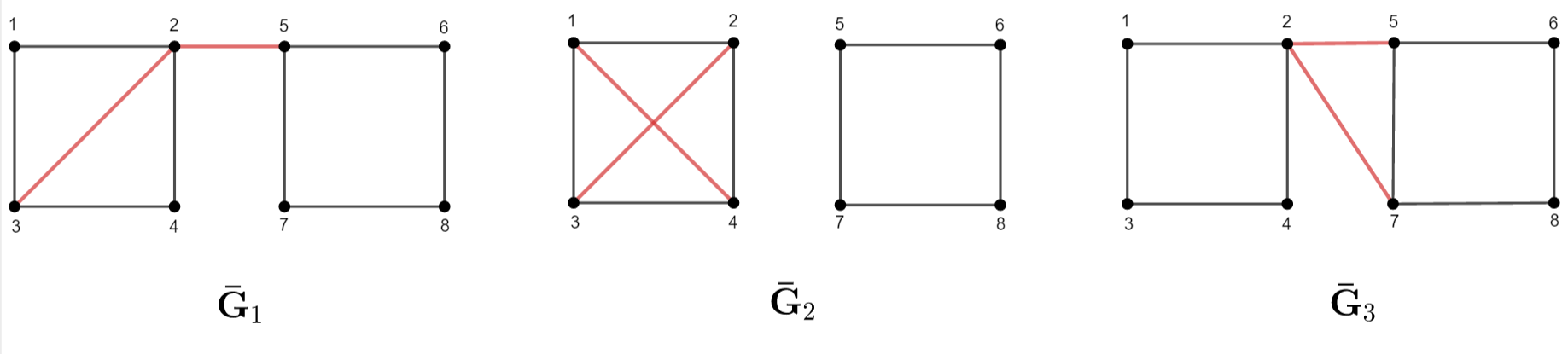}
    \caption{Adding two links between two separated networks}
    \label{fg:two_links_add}
    \end{figure}
    
    \begin{table}[htp]
\caption{Aggregate Katz-Bonacich centralities ($\protect\delta =0.21$)}
\begin{center}
\begin{tabular}{cc|cc}
\hline\hline
Add one link & $b\left(\right) $ & Add two links & $b\left( \right) $ \\ \hline
$\hat{\mathbf{G}}_{1}$ & 15.4198 & $\bar{\mathbf{G}}_{1}$ & 17.7010 \\ 
$\hat{\mathbf{G}}_{2}$ & 15.4689* & $\bar{\mathbf{G}}_{2}$ & 17.7074 \\ 
&  & $\bar{\mathbf{G}}_{3}$ & 17.7547* \\ \hline
\end{tabular}\end{center}
\label{tab:centrality_comparison}
\end{table}
   \end{itemize}
\end{ex}

\section{Extensions and concluding remarks}

\label{extensions}

\subsection{Alternative network models}

\label{sec:5.1}

Our analysis so far focuses on the impact of structural interventions on
the Katz-Bonacich centralities in the baseline model of \cite{Ballester2006}%
. Since Katz-Bonacich centrality plays a critical role for many network
models, our results (and subsequent applications) naturally extend to
these alternative models. For instance, \cite{CURRARINI201740} extend the
single-activity network model of \cite{Ballester2006} with direct
complements to incorporate indirect substitutes among players with distance
two, and characterize the equilibrium using both $\mathbf{G}$ and $\mathbf{G}%
^2$. In the Appendix, we show that the equilibrium in \cite{CURRARINI201740}
can be written as a linear combination of two Katz-Bonacich centralities. 
\cite{Ballester2006} extend the baseline model in equation \eqref{eq:complements} to
allow global substitution and show that the aggregate action is a monotone
transformation of Katz-Bonacich centralities. In addition, \cite{D2x}
consider a network game with multiple activities, and show that the
equilibrium can be represented as the weights sum of two Katz-Bonacich
centralities (with different synergy parameters and characteristics). See
Appendix \ref{app-alt-C} for details.

\subsection{Hybrid interventions}

\label{sec:hybrid}

We can study the effect of general interventions that combine both structural
and characteristic interventions in networks. A general hybrid intervention
is given by $\left( \mathbf{C},\Delta \boldsymbol{\theta }\right) $, where $%
\mathbf{C}$ is the structural intervention and $\Delta \boldsymbol{\theta }$
 the characteristic intervention. The hybrid intervention $\left( \mathbf{C%
},\Delta \boldsymbol{\theta }\right) $ on network game $\Gamma \left( 
\mathbf{G},\boldsymbol{\theta }\right) $ can be viewed as a structural
intervention $\mathbf{C}$ on network game $\Gamma \left( \mathbf{G},%
\boldsymbol{\theta +}\Delta \boldsymbol{\theta }\right) $. By Proposition %
\ref{prop-1}, this hybrid intervention $\left( \mathbf{C},\Delta \boldsymbol{%
\theta }\right) $ is outcome equivalent to a characteristic intervention 
\begin{equation*}
\widetilde{\Delta \boldsymbol{\theta }^{\ast }}=\Delta \boldsymbol{\theta }+%
\begin{bmatrix}
\mathbf{0} \\ 
\delta \mathbf{C}_{SS}\left( \mathbf{I}-\delta \mathbf{M}_{SS}\left( \mathbf{%
G}\right) \mathbf{C}_{SS}\right) ^{-1}\mathbf{b}_{S}\left( \mathbf{G},%
\boldsymbol{\theta }+\Delta \boldsymbol{\theta }\right)%
\end{bmatrix}%
\end{equation*}%
of $\Gamma\left( \mathbf{G},\boldsymbol{\theta }\right) $. Thus, the new
equilibrium after this hybrid intervention is given by%
\begin{equation*}
\mathbf{\hat{x}}^{\ast }=\mathbf{b}\left( \mathbf{G},\boldsymbol{\theta }%
\right) +\mathbf{M}\left( \mathbf{G}\right) \widetilde{\Delta \boldsymbol{%
\theta }^{\ast }}.
\end{equation*}%
This characterization of the equilibrium effects of hybrid interventions enables
us to study optimal combinations of intervention policies in networks.

\subsection{Concluding remarks}

\label{Discussion}

In this paper, we present a theory of interventions in network. By showing
an equivalence between a structural intervention and an \emph{endogenously determined}
characteristic intervention, we analyze how these two types of interventions
affect the equilibrium actions and offer new insights regarding the optimal
interventions in a range of applications.

We discuss several venues for future work. First, this paper mainly focuses
on the benefit of structural interventions without explicitly modeling the
cost of cutting/building links and nodes. It would be interesting to study the
optimal intervention policy with a budget for the cost of interventions
(see \citet*{Galeotti2017}).  Second, we treat two instruments -- i.e.,
characteristics and social links -- independently in our analysis. In some
contexts, intervention in one space (e.g., the characteristics) may induce
endogenous responses in the other space (the network links). For instance, %
\citet*{banerjee2018} show that new links are formed and existing links are
removed after exposure to formal credit markets.\footnote{See \cite{cabrales2011social} and \cite{golub2021games} for network games with endogenous link formation.}  Extending hybrid
interventions to accommodate interdependence between two instruments is 
an intriguing subject.  
Third, our analysis mainly focuses on the effects of interventions on the aggregate action.
 It would be  interesting to  explore the distributional effects (such as inequality) of actions.
Finally, It would be natural to extend our approach to network games with nonlinear responses (see, for instance, \cite%
{Allouch2015,Ben2018,ZZ2021}). These and other generalizations will
enrich our understanding of optimal interventions in economic settings that
involve networks.


%
 \newpage

 \appendix

\begin{center}
\textbf{\Large Appendix }
\end{center}


\section{Proofs}

\noindent \textbf{Proof of Lemma \ref{Partial}:} In the game $\Gamma \left( 
\mathbf{G,\theta }\right) $ the equilibrium action profile is $\mathbf{x}%
^{\ast }=\mathbf{M}\left( \mathbf{G}\right) \boldsymbol{\theta }$, and the
aggregate equilibrium action is ${x}^{\ast }=\mathbf{1}^{\prime }\mathbf{M}%
\left( \mathbf{G}\right) \boldsymbol{\theta }=\mathbf{b}^{\prime }(\mathbf{G}%
)\boldsymbol{\theta }$. Since the network $\mathbf{G}$ is fixed for a
characteristic intervention, the results directly follow. \hfill $\Box $

\medskip


\noindent \textbf{Proofs of Lemma \ref{lm-2}:} The equilibrium actions of $%
\Gamma \left( \mathbf{G},\boldsymbol{\theta }\right) $ satisfy $\mathbf{x}%
^{\ast }=\boldsymbol{\theta }+\delta \mathbf{Gx}^{\ast }$. Under the
structural intervention $\mathbf{C}$, the new equilibrium action profile $%
\mathbf{\hat{x}}^{\ast }$ satisfies 
\begin{equation*}
\mathbf{\hat{x}}^{\ast }=\boldsymbol{\theta }+\delta \left( \mathbf{G}+%
\mathbf{C}\right) \mathbf{\hat{x}}^{\ast }=\left( \boldsymbol{\theta }+%
\underbrace{\delta \mathbf{C\hat{x}}^{\ast }}_{=\mathbf{\Delta }\boldsymbol{%
\theta }^{\ast }}\right) +\delta \mathbf{G\hat{x}}^{\ast }\text{.}
\end{equation*}%
That is, the structural intervention $\mathbf{C}$ is outcome equivalent to a
change of players' intrinsic marginal utilities from $\boldsymbol{\theta }$
to $\boldsymbol{\theta }+\mathbf{\Delta }\boldsymbol{\theta }^{\ast }=%
\boldsymbol{\theta }+\delta \mathbf{C\hat{x}}^{\ast }$. Given $\mathbf{C}=%
\begin{bmatrix}
\mathbf{0} & \mathbf{0} \\ 
\mathbf{0} & \mathbf{C}_{SS}%
\end{bmatrix}%
$, we obtain 
\begin{equation*}
\mathbf{\Delta }\boldsymbol{\theta }^{\ast }=\delta \mathbf{C\hat{x}}^{\ast
}=\delta 
\begin{bmatrix}
\mathbf{0} & \mathbf{0} \\ 
\mathbf{0} & \mathbf{C}_{SS}%
\end{bmatrix}%
\begin{bmatrix}
\mathbf{\hat{x}}_{S^{C}}^{\ast } \\ 
\mathbf{\hat{x}}_{S}^{\ast }%
\end{bmatrix}%
=%
\begin{bmatrix}
\mathbf{0} \\ 
\delta \mathbf{C}_{SS}\mathbf{\hat{x}}_{S}^{\ast }%
\end{bmatrix}%
\text{.}
\end{equation*}%
That is, the structural intervention $\mathbf{C}$ is outcome equivalence to
changing the characteristics of players in $S$ by $\mathbf{\Delta }%
\boldsymbol{\theta }_{S}^{\ast }=\delta \mathbf{C}_{SS}\mathbf{\hat{x}}%
_{S}^{\ast }$. Moreover, from equation \eqref{eq:partial}, $\mathbf{\hat{x}}%
_{S}^{\ast }$ must satisfy the following: 
\begin{equation*}
\mathbf{\hat{x}}_{S}^{\ast }=\mathbf{b}_{S}\left( \mathbf{G},\boldsymbol{%
\theta }\right) +\mathbf{M}_{SS}\left( \mathbf{G}\right) \mathbf{\Delta }%
\boldsymbol{\theta }_{S}^{\ast }=\mathbf{b}_{S}\left( \mathbf{G},\boldsymbol{%
\theta }\right) +\delta \mathbf{M}_{SS}\left( \mathbf{G}\right) \mathbf{C}%
_{SS}\mathbf{\hat{x}}_{S}^{\ast }\text{.}
\end{equation*}%
Solving it yields $\mathbf{\hat{x}}_{S}^{\ast }=\left( \mathbf{I}-\delta 
\mathbf{M}_{SS}\left( \mathbf{G}\right) \mathbf{C}_{SS}\right) ^{-1}\mathbf{b%
}_{S}\left( \mathbf{G},\boldsymbol{\theta }\right). $ Consequently, we
obtain 
\begin{equation*}
\mathbf{\Delta }\boldsymbol{\theta }_{S}^{\ast }=\delta \mathbf{C}_{SS}%
\mathbf{\hat{x}}_{S}^{\ast }=\delta \mathbf{C}_{SS}\left( \mathbf{I}-\delta 
\mathbf{M}_{SS}\left( \mathbf{G}\right) \mathbf{C}_{SS}\right) ^{-1}\mathbf{b%
}_{S}\left( \mathbf{G},\boldsymbol{\theta }\right).
\end{equation*}
\hfill $\square $ 

\medskip

\noindent \textbf{Proof of Proposition \ref{prop-1}:} It directly follows
from Lemmas \ref{lm-1} and \ref{lm-2}. \hfill $\square $ 

\medskip

\noindent \textbf{Proof of Corollary \ref{cor-1}:} Define $\eta (t)=\mathbf{1%
}^{\prime }(\mathbf{I}-\delta (\mathbf{G}+t\mathbf{C}))^{-1}\mathbf{1},t\in
\lbrack 0,1]$. Since we have assumed that $\mathbf{I}-\delta \mathbf{G}$ and 
$\mathbf{I}-\delta (\mathbf{G}+\mathbf{C})$ are both symmetric positive
definite, $(\mathbf{I}-\delta (\mathbf{G}+t\mathbf{C}))$ is positive
definite for any $t\in \lbrack 0,1]$, and hence $\eta $ is welldefined. Given $%
\boldsymbol{\theta }=\mathbf{1}$, 
\begin{equation*}
\eta (0)=\mathbf{1}^{\prime }(\mathbf{I}-\delta (\mathbf{G})^{-1}\mathbf{1}%
=b(\mathbf{G},\mathbf{1})
\end{equation*}%
is the equilibrium aggregate action before the intervention, and 
\begin{equation*}
\eta (1)=\mathbf{1}^{\prime }(\mathbf{I}-\delta (\mathbf{G}+\mathbf{C})^{-1}%
\mathbf{1}=b(\mathbf{G+C},\mathbf{1})
\end{equation*}%
is the equilibrium aggregate action after the intervention. Direct
computation shows that\footnote{%
We use the fact that $d(A^{-1})=-A^{-1}(dA)A^{-1}$.} 
\begin{equation*}
\begin{split}
\eta ^{\prime }(0)=\eta ^{\prime }(t)|_{t=0}& =\mathbf{1}^{\prime }(\mathbf{I%
}-\delta (\mathbf{G}+t\mathbf{C}))^{-1}\delta \mathbf{C}(\mathbf{I}-\delta (%
\mathbf{G}+t\mathbf{C}))^{-1}\mathbf{1}|_{t=0} \\
& =\mathbf{1}^{\prime }(\mathbf{I}-\delta \mathbf{G})^{-1}\delta \mathbf{C}(%
\mathbf{I}-\delta \mathbf{G})^{-1}\mathbf{1}=\delta \mathbf{b}^{\prime }(%
\mathbf{G})\mathbf{C}\mathbf{b}(\mathbf{G}).
\end{split}%
\end{equation*}
Critically, $\eta (\cdot )$ is convex in $t$ by Lemma \ref{lm:cvx} below;
therefore, $\eta (1)-\eta (0)\geq \eta ^{\prime }(0)(1-0).$ In other words, $%
b(\mathbf{G+C},\mathbf{1})-b(\mathbf{G},\mathbf{1})\geq \delta \mathbf{b}%
^{\prime }(\mathbf{G})\mathbf{C}\mathbf{b}(\mathbf{G}), $ which implies
Corollary \ref{cor-1}.\footnote{%
As seen from the proof, Corollary \ref{cor-1} holds for weighted undirected
networks as well.} \hfill $\Box $ 

\medskip

\begin{lem}
\label{lm:cvx} Let $\mathcal{O}$ denote the set of $n$ by $n$ symmetric
positive definite matrices. Then the function $V(\mathbf{A}):=\mathbf{1}%
^{\prime }\mathbf{A}^{-1}\mathbf{1}$ is convex in $\mathbf{A}\in\mathcal{O}$.
\end{lem}

\noindent \textbf{Proof of Lemma \ref{lm:cvx}:} Define $H(\mathbf{A},\mathbf{%
x})=2\mathbf{1}^{\prime }x-\mathbf{x}^{\prime }\mathbf{A}\mathbf{x}$, where $%
\mathbf{A}\in \mathcal{O},\mathbf{x}\in \mathbf{R}^{n}$. Fixing a positive
definite matrix $\mathbf{A}\in \mathcal{O}$, $H(\mathbf{A},\cdot )$ is
strictly concave in $x$ with the maximum value 
\begin{equation*}
\max_{\mathbf{x}\in \mathbf{R}^{n}}H(\mathbf{A},\mathbf{x})=\mathbf{1}%
^{\prime }\mathbf{A}^{-1}\mathbf{1}=V(\mathbf{A}),
\end{equation*}
obtained at $\mathbf{x}^{\ast }=\mathbf{A}^{-1}\mathbf{1}$. Moreover, $H(%
\mathbf{A},\mathbf{x})$ is linear in $\mathbf{A}$ for fixed $\mathbf{x}$, so 
$V(\mathbf{A})=\max_{\mathbf{x}\in \mathbf{R}^{n}}H(\mathbf{A},\mathbf{x})$
is convex in $\mathbf{A}\in \mathcal{O}$, since the maximum of a family of
linear functions is convex (see \cite{boyd2004convex}). \hfill $\Box $ 

\medskip

\noindent \textbf{Proof of Lemma \ref{key group}:} As demonstrated in the
main text, $d_{S}\left( \mathbf{G},\boldsymbol{\theta }\right) $ exactly
equals the effect of the characteristic intervention $\Delta \boldsymbol{%
\theta }_{S}=\left( \mathbf{M}_{SS}\left( \mathbf{G}\right) \right) ^{-1}%
\mathbf{b}_{S}\left( \mathbf{G},\boldsymbol{\theta }\right) $ on the
aggregate action. Therefore, by Lemma \ref{Partial}, $d_{S}\left( \mathbf{G},%
\boldsymbol{\theta }\right) =\mathbf{b}_{S}^{\prime }\left( \mathbf{G}%
\right) \Delta \boldsymbol{\theta }_{S}=\mathbf{b}_{S}^{\prime }\left( 
\mathbf{G}\right) \left( \mathbf{M}_{SS}\left( \mathbf{G}\right) \right)
^{-1}\mathbf{b}_{S}\left( \mathbf{G},\boldsymbol{\theta }\right) $. \hfill $%
\Box $ 

\medskip

\noindent\textbf{Proof of Proposition \ref{dominated groups}:} Part (i) is
obvious. For part (ii), we first define $\mathbf{v}:=\left( \mathbf{M}%
_{SS}\left( \mathbf{G}\right) \right) ^{-1}\mathbf{b}_{S}\left( \mathbf{G},%
\boldsymbol{1}\right) $. We first show that $\mathbf{v}$ is a positive
vector, i.e, $\mathbf{v}\succeq \mathbf{0}$: 
\begin{eqnarray*}
\mathbf{v} &=&\left( \mathbf{M}_{SS}\left( \mathbf{G}\right) \right) ^{-1}%
\mathbf{b}_{S}\left( \mathbf{G},\boldsymbol{1}\right) =\left( \mathbf{M}%
_{SS}\left( \mathbf{G}\right) \right) ^{-1}\left( \mathbf{M}_{SS^{C}}\left( 
\mathbf{G}\right) \mathbf{1}_{|S^{C}|}+\mathbf{M}_{SS}\left( \mathbf{G}%
\right) \mathbf{1}_{|S|}\right) \\
&=&\left( \mathbf{M}_{SS}\left( \mathbf{G}\right) \right) ^{-1}\mathbf{M}%
_{SS^{C}}\left( \mathbf{G}\right) \mathbf{1}_{|S^{C}|}+\mathbf{1}_{|S|} = 
\underbrace{\delta \mathbf{G}_{SS^{C}}\left( \mathbf{I}-\delta \mathbf{G}%
_{S^{C}S^{C}}\right) ^{-1}\mathbf{1}_{|S^{C}|}}_{\succeq \mathbf{0}}+\mathbf{%
1}_{|S|}\succeq \mathbf{0},
\end{eqnarray*}%
where in the last equality we use the identity $\left( \mathbf{M}_{SS}\left( 
\mathbf{G}\right) \right) ^{-1}\mathbf{M}_{SS^{C}}=\delta \mathbf{G}%
_{SS^{C}}\left( \mathbf{I}-\delta \mathbf{G}_{S^{C}S^{C}}\right) ^{-1}$.
Moreover, given $\mathbf{b}_{S}\left( \mathbf{G}\right) \preceq \mathbf{b}%
_{S^{\prime }}\left( \mathbf{G}\right) $,  $\mathbf{M}_{SS}\left( \mathbf{%
G}\right) \succeq \mathbf{M}_{S^{\prime }S^{\prime }}\left( \mathbf{G}%
\right) $, and $\mathbf{v}\succeq \mathbf{0}$, we have 
\begin{equation}
d_{S}\left( \mathbf{G},\mathbf{1}\right) =2\mathbf{b}_{S}\left( \mathbf{G}%
\right) \mathbf{v}-\mathbf{v}^{\prime }\mathbf{M}_{SS}\left( \mathbf{G}%
\right) \mathbf{v}\leq 2\mathbf{b}_{S^{\prime }}^{\prime }\left( \mathbf{G}%
\right) \mathbf{v}-\mathbf{v}^{\prime }\mathbf{M}_{S^{\prime }S^{\prime
}}\left( \mathbf{G}\right) \mathbf{v}.  \label{eq-pr2-1}
\end{equation}%
Solving the following concave programming yields\footnote{%
As a principle submatrix of positive definite matrix $\left( \mathbf{I}%
-\delta \mathbf{G}\right) ^{-1} $, $\mathbf{M}_{S^{\prime }S^{\prime
}}\left( \mathbf{G}\right) $ is also positive definite.} 
\begin{equation}
\underset{\mathbf{x\in }\mathbb{R}^{|S|}}{\max }\left\{ 2\mathbf{b}%
_{S^{\prime }}^{\prime }\left( \mathbf{G}\right) \mathbf{x}-\mathbf{x}%
^{\prime }\mathbf{M}_{S^{\prime }S^{\prime }}\left( \mathbf{G}\right) 
\mathbf{x}\right\} =\mathbf{b}_{S^{\prime }}^{\prime }\left( \mathbf{G}%
\right) \left( \mathbf{M}_{S^{\prime }S^{\prime }}\left( \mathbf{G}\right)
\right) ^{-1}\mathbf{b}_{S^{\prime }}\left( \mathbf{G}\right) =d_{S^{\prime
}}\left( \mathbf{G},\mathbf{1}\right) .  \label{eq-pr2-2}
\end{equation}%
By optimality, 
\begin{equation}
\underset{\mathbf{x\in }\mathbb{R}^{|S|}}{\max }\left\{ 2\mathbf{b}%
_{S^{\prime }}^{\prime }\left( \mathbf{G}\right) \mathbf{x}-\mathbf{x}%
^{\prime }\mathbf{M}_{S^{\prime }S^{\prime }}\left( \mathbf{G}\right) 
\mathbf{x}\right\} \geq 2\mathbf{b}_{S^{\prime }}^{\prime }\left( \mathbf{G}%
\right) \mathbf{v}-\mathbf{v}^{\prime }\mathbf{M}_{S^{\prime }S^{\prime
}}\left( \mathbf{G}\right) \mathbf{v}.  \label{eq-pr2-3}
\end{equation}%
Combining equations \eqref{eq-pr2-1}, \eqref{eq-pr2-2}, and \eqref{eq-pr2-3} yields $%
d_{S}\left( \mathbf{G},\mathbf{1}\right) \leq d_{S^{\prime }}\left( \mathbf{G%
},\mathbf{1}\right) $. \hfill $\square $ 
\medskip




\noindent \textbf{Proof of Proposition \ref{prop-3}:}

\begin{remark}
\label{rm-W} We can have alternative expressions for blocks of the matrix $%
\mathbf{W}$, as follows: 
\begin{eqnarray*}
\mathbf{W}_{S^{C}S^{C}}\left( \mathbf{G},S\right) &=&\left( \mathbf{I}%
-\delta \mathbf{G}_{S^{C}S^{C}}\right) ^{-1}\text{;} \\
\mathbf{W}_{S^{C}S}\left( \mathbf{G},S\right) &=&\left( \mathbf{I}-\delta 
\mathbf{G}_{S^{C}S^{C}}\right) ^{-1}\delta \mathbf{G}_{S^{C}S}\text{;} \\
\mathbf{W}_{SS}\left( \mathbf{G},S\right) &=&\delta \mathbf{G}%
_{SS^{C}}\left( \mathbf{I}-\delta \mathbf{G}_{S^{C}S^{C}}\right) ^{-1}\delta 
\mathbf{G}_{S^{C}S}+\delta \mathbf{G}_{SS}+\mathbf{I}\text{.}
\end{eqnarray*}%
These expressions directly follow from the definition of $\mathbf{W}$.
Unlike Proposition \ref{prop-3}, these expressions use the
centralities in the remaining network $\mathbf{G}_{S^{C}S^{C}}$.
\end{remark}

The Leontief inverse matrix can be written in block form%
\begin{equation*}
\left( \mathbf{I}-\delta \mathbf{G}\right) ^{-1}=\left[ 
\begin{array}{cc}
\mathbf{I}-\delta \mathbf{G}_{S^{C}S^{C}} & -\delta \mathbf{G}_{S^{C}S} \\ 
-\delta \mathbf{G}_{SS^{C}} & \mathbf{I}-\delta \mathbf{G}_{SS}%
\end{array}%
\right] ^{-1}=\left[ 
\begin{array}{cc}
\mathbf{M}_{S^{C}S^{C}}\left( \mathbf{G}\right) & \mathbf{M}_{S^{C}S}\left( 
\mathbf{G}\right) \\ 
\mathbf{M}_{SS^{C}}\left( \mathbf{G}\right) & \mathbf{M}_{SS}\left( \mathbf{G%
}\right)%
\end{array}%
\right] .
\end{equation*}%
For easy notation, let $\mathbf{I}-\delta \mathbf{G}_{S^{C}S^{C}}=\mathbf{A}$%
, $-\delta \mathbf{G}_{S^{C}S}=\mathbf{B}$, $-\delta \mathbf{G}_{SS^{C}}=%
\mathbf{C}$ and $\mathbf{I}-\delta \mathbf{G}_{SS}=\mathbf{D}$. Then by the
remark above, we have $\mathbf{W}_{S^{C}S^{C}}\left( \mathbf{G},S\right) =%
\mathbf{A}^{-1}$, $\mathbf{W}_{S^{C}S}\left( \mathbf{G},S\right) =-\mathbf{A}%
^{-1}\mathbf{B}$, and $\mathbf{W}_{SS}\left( \mathbf{G},S\right) =\mathbf{CA}%
^{-1}\mathbf{B-D+}2\mathbf{I}$. Using the block matrix inversion,%
\begin{eqnarray*}
\left[ 
\begin{array}{cc}
\mathbf{A} & \mathbf{B} \\ 
\mathbf{C} & \mathbf{D}%
\end{array}%
\right] ^{-1} &=&\left[ 
\begin{array}{cc}
\mathbf{A}^{-1}\mathbf{+A}^{-1}\mathbf{B}\left( \mathbf{D}-\mathbf{CA}^{-1}%
\mathbf{B}\right) ^{-1}\mathbf{CA}^{-1} & -\mathbf{A}^{-1}\mathbf{B}\left( 
\mathbf{D}-\mathbf{CA}^{-1}\mathbf{B}\right) ^{-1} \\ 
\mathbf{-\left( \mathbf{D}-\mathbf{CA}^{-1}\mathbf{B}\right) ^{-1}CA}^{-1} & 
\left( \mathbf{D}-\mathbf{CA}^{-1}\mathbf{B}\right) ^{-1}%
\end{array}%
\right] \\
&=&\left[ 
\begin{array}{cc}
\mathbf{M}_{S^{C}S^{C}}\left( \mathbf{G}\right) & \mathbf{M}_{S^{C}S}\left( 
\mathbf{G}\right) \\ 
\mathbf{M}_{SS^{C}}\left( \mathbf{G}\right) & \mathbf{M}_{SS}\left( \mathbf{G%
}\right)%
\end{array}%
\right] \text{.}
\end{eqnarray*}%
Thus, $\left( \mathbf{D}-\mathbf{CA}^{-1}\mathbf{B}\right) ^{-1}=\mathbf{M}%
_{SS}\left( \mathbf{G}\right) $, which implies 
\begin{equation*}
\mathbf{W}_{SS}\left( \mathbf{G},S\right) =\mathbf{CA}^{-1}\mathbf{B-D+}2%
\mathbf{I}=2\mathbf{I+}\left( \mathbf{M}_{SS}\left( \mathbf{G}\right)
\right) ^{-1}\text{.}
\end{equation*}%
We further have $-\mathbf{A}^{-1}\mathbf{B}%
\left( \mathbf{D}-\mathbf{CA}^{-1}\mathbf{B}\right) ^{-1}=\mathbf{M}%
_{S^{C}S}\left( \mathbf{G}\right) $. Therefore, 
\begin{equation*}
\mathbf{A}^{-1}\mathbf{B=\mathbf{W}_{S^{C}S}\left( \mathbf{G},S\right) =-M}%
_{S^{C}S}\left( \mathbf{G}\right) \left( \mathbf{M}_{SS}\left( \mathbf{G}%
\right) \right) ^{-1}\text{.}
\end{equation*}%
In addition, $\mathbf{A}^{-1}\mathbf{B}\left( \mathbf{D}-\mathbf{CA}^{-1}%
\mathbf{B}\right) ^{-1}\mathbf{CA}^{-1}=\mathbf{M}_{S^{C}S}\left( \mathbf{G}%
\right) \left( \mathbf{M}_{SS}\left( \mathbf{G}\right) \right) ^{-1}\mathbf{M%
}_{SS^{C}}\left( \mathbf{G}\right) $. Substituting in the identity $\mathbf{A%
}^{-1}\mathbf{+A}^{-1}\mathbf{B}\left( \mathbf{D}-\mathbf{CA}^{-1}\mathbf{B}%
\right) ^{-1}\mathbf{CA}^{-1}=\mathbf{M}_{S^{C}S^{C}}\left( \mathbf{G}%
\right) $, we can get 
\begin{equation*}
\mathbf{A}^{-1}=\mathbf{W}_{S^{C}S^{C}}\left( \mathbf{G},S\right) =\mathbf{M}%
_{S^{C}S^{C}}\left( \mathbf{G}\right) -\mathbf{M}_{S^{C}S}\left( \mathbf{G}%
\right) \left( \mathbf{M}_{SS}\left( \mathbf{G}\right) \right) ^{-1}\mathbf{M%
}_{SS^{C}}\left( \mathbf{G}\right) \text{.}
\end{equation*}

\medskip Let $S=A\cup B$. Then the inverse of matrix $\boldsymbol{M}_{SS}$
is given as {\small \ 
\begin{equation*}
\begin{aligned}(\boldsymbol{M}_{SS})^{-1}= & \left[\begin{array}{cc}
\boldsymbol{M}_{AA} & \boldsymbol{M}_{AB}\\ \boldsymbol{M}_{BA} &
\boldsymbol{M}_{BB} \end{array}\right]^{-1}\\ = & \left[\begin{array}{cc}
(\boldsymbol{M}_{AA})^{-1}\left(\boldsymbol{I}+\boldsymbol{M}_{AB}\left(%
\boldsymbol{W}_{BB}(\boldsymbol{G},A)\right)^{-1}\boldsymbol{M}_{BA}(%
\boldsymbol{M}_{AA})^{-1}\right) &
-(\boldsymbol{M}_{AA})^{-1}\boldsymbol{M}_{AB}\left(\boldsymbol{W}_{BB}(%
\boldsymbol{G},A)\right)^{-1}\\
-\left(\boldsymbol{W}_{BB}(\boldsymbol{G},A)\right)^{-1}\boldsymbol{M}_{BA}(%
\boldsymbol{M}_{AA})^{-1} &
\left(\boldsymbol{W}_{BB}(\boldsymbol{G},A)\right)^{-1} \end{array}\right]\\
= & \left[\begin{array}{cc}
\left(\boldsymbol{W}_{AA}(\boldsymbol{G},B)\right)^{-1} &
-\left(\boldsymbol{W}_{AA}(\boldsymbol{G},B)\right)^{-1}\boldsymbol{M}_{AB}(%
\boldsymbol{M}_{BB})^{-1}\\
-(\boldsymbol{M}_{BB})^{-1}\boldsymbol{M}_{BA}\left(\boldsymbol{W}_{AA}(%
\boldsymbol{G},B)\right)^{-1} &
(\boldsymbol{M}_{BB})^{-1}\left(\boldsymbol{I}+\boldsymbol{M}_{BA}\left(%
\boldsymbol{W}_{AA}(\boldsymbol{G},B)\right)^{-1}\boldsymbol{M}_{AB}(%
\boldsymbol{M}_{BB})^{-1}\right) \end{array}\right] \end{aligned} .
\end{equation*}
}

Using \eqref{eq: Lemma 1 extension 3}, we obtain that
\begin{equation*}
\begin{aligned}\boldsymbol{W}_{AB}(\boldsymbol{G},A\cup B) &
=\left(2\boldsymbol{I}-(\boldsymbol{M}_{SS})^{-1}\right)_{AB}=-\left((%
\boldsymbol{M}_{SS})^{-1}\right)_{AB}\\ &
=(\boldsymbol{M}_{AA})^{-1}\boldsymbol{M}_{AB}\left(\boldsymbol{W}_{BB}(%
\boldsymbol{G},A)\right)^{-1}\\ &
=\left(\boldsymbol{W}_{AA}(\boldsymbol{G},B)\right)^{-1}\boldsymbol{M}_{AB}(%
\boldsymbol{M}_{BB})^{-1}. \end{aligned}
\end{equation*}

\medskip

\noindent \textbf{Proof of Proposition \ref{pro-bridge}:} For Proposition \ref%
{pro-bridge}, it suffices to show equation \eqref{plus-ij}. Indeed, when the bridge
link between $i\in N_{1}$ and $j\in N_{2}$ is added, the new equilibrium
efforts of $i$ and $j$ satisfy%
\begin{equation}  \label{bridge-xij}
\begin{cases}
\hat{x}_{i}^{\ast }{{=b_{i}\left( \mathbf{N}^{1}\right) +m_{ii}\left( 
\mathbf{N}^{1}\right) \underbrace{\delta \hat{x}_{j}^{\ast }}_{=\Delta
\theta _{i}^{\ast }}}} \\ 
\hat{x}_{j}^{\ast }{{=b_{j}\left( \mathbf{N}^{2}\right) +m_{jj}\left( 
\mathbf{N}^{2}\right) \underbrace{\delta \hat{x}_{i}^{\ast }}_{=\Delta
\theta _{j}^{\ast }}}}%
\end{cases}%
\text{.}
\end{equation}%
Here we have translated the structural intervention (adding the link $i-j$)
into the corresponding characteristic intervention: $\Delta \theta
_{i}^{\ast }=\delta \hat{x}_{j}^{\ast }$, $\Delta \theta _{j}^{\ast }=\delta 
\hat{x}_{i}^{\ast }$ and $\Delta \theta _{k}^{\ast }=0$ for all $k\notin
\left\{ i,j\right\} $ (see Lemma \ref{lm-2}). Note that $m_{ij}=m_{ji}=0$ since
two networks are initially isolated. Simple algebra yields 
\begin{eqnarray*}
\hat{x}_{i}^{\ast } &=&\frac{b_{i}\left( \mathbf{N}^{1},\boldsymbol{\theta }%
^{1}\right) +\delta m_{ii}\left( \mathbf{N}^{1}\right) b_{j}\left( \mathbf{N}%
^{2},\boldsymbol{\theta }^{2}\right) }{1-\delta ^{2}m_{ii}\left( \mathbf{N}%
^{1}\right) m_{jj}\left( \mathbf{N}^{2}\right) }\text{,~~~} \hat{x}%
_{j}^{\ast } =\frac{b_{j}\left( \mathbf{N}^{2},\boldsymbol{\theta }%
^{2}\right) +\delta m_{jj}\left( \mathbf{N}^{2}\right) b_{i}\left( \mathbf{N}%
^{1},\boldsymbol{\theta }^{1}\right) }{1-\delta ^{2}m_{ii}\left( \mathbf{N}%
^{1}\right) m_{jj}\left( \mathbf{N}^{2}\right) }\text{.}
\end{eqnarray*}%
By Proposition \ref{prop-1}, the change in aggregate action equals 
\begin{equation*}
b\left( \mathbf{G+E}_{ij}\right)-b\left( \mathbf{G}\right)=b_{i}\left( 
\mathbf{N}^{1}\right) \underbrace{\delta \hat{x}_{j}^{\ast }}_{=\Delta 
\boldsymbol{\theta }_{i}^{\ast }}+b_{j}\left( \mathbf{N}^{2}\right) {{%
\underbrace{\delta \hat{x}_{i}^{\ast }}_{=\Delta \boldsymbol{\theta }%
_{j}^{\ast }}}}=\delta L_{ij}\left( \mathbf{N}^{1},\mathbf{N}^{2}\right).
\end{equation*}

\medskip

\noindent \textbf{Proof of Corollary \ref{bridge cor}:} For item (i), we
first note that by Definition \ref{Bridge index}, $L_{ij}$ clearly increases
with $b_{i}$ and $m_{ii}$ for each given $j$. The claim just follows.

For item (ii), given $m_{jj}=m_{j^{\prime }j^{\prime }}$ and $b_j\geq b_{j'}$, we have 
\begin{eqnarray*}
L_{ij}-L_{ij^{\prime }} &=&\frac{\delta m_{ii}\left( b_{j}^{2}-b_{j^{\prime
}}^{2}\right) }{1-\delta ^{2}m_{ii}m_{jj}}+\frac{2b_{i}\left(
b_{j}-b_{j^{\prime }}\right) }{1-\delta ^{2}m_{ii}m_{jj}} = \frac{\delta \left( b_{j}^{2}-b_{j^{\prime }}^{2}\right) }{\frac{1}{m_{ii}%
}-\delta ^{2}m_{jj}}+\frac{2b_{i}\left( b_{j}-b_{j^{\prime }}\right) }{%
1-\delta ^{2}m_{ii}m_{jj}}\text{,}
\end{eqnarray*}
which clearly increases in $m_{ii}$.
The result just follows by noting that $b_i=b_{i'}$ and $m_{ii}\geq m_{i'i'}$.

For item (iii), we apply the Taylor expansions to obtain that 
\begin{equation*}
b_{k}\left( \mathbf{N}^{1}\right) =1+\delta e_{k}\left( \mathbf{N}%
^{1}\right) +O\left( \delta ^{2}\right) ,m_{kk}\left( \mathbf{N}^{1}\right)
=1+O\left( \delta ^{2}\right) ,~~k\in N_{1},
\end{equation*}%
where $O\left( \delta ^{2}\right) $ denotes a real-valued function such that 
$\lim \sup_{\delta \rightarrow 0}|\frac{\mathcal{O}\left( \delta ^{2}\right) 
}{\delta ^{2}}|<\infty $. Consequently, 
\begin{equation*}
L_{ij}\left( \mathbf{N}^{1},\mathbf{N}^{2}\right)= 2+2\delta \left(1+ e_{i}\left( \mathbf{N%
}^{1}\right) +e_{j}\left( \mathbf{N}^{2}\right) \right) +\mathcal{O}\left(
\delta ^{2}\right) \text{.}
\end{equation*}%
Thus, when $\delta $ is sufficiently small, only the degree centrality
matters for the bridge index $L_{ij}$. \hfill $\square $ 
\medskip

\noindent\textbf{Proof of Lemma \ref{lm-keylink}:} The proof is similar to
that of Proposition \ref{pro-bridge}, with the exception that $m_{ij}$ is not
necessarily zero. For item (i), applying Proposition \ref{prop-1} with $%
S=\{i,j\}$ and $\mathbf{C}_{SS}=%
\begin{bmatrix}
0 & 1 \\ 
1 & 0%
\end{bmatrix}%
$ yields 
\begin{equation*}
\Delta x^*=\mathbf{b}_{S}^{\prime }\left( \mathbf{G}\right) \Delta 
\boldsymbol{\theta }_{S}^{\ast } =\mathbf{b}_{S}^{\prime } \delta \mathbf{C}%
_{SS}\left( \mathbf{I}-\delta \mathbf{M}_{SS}\left( \mathbf{G}\right) 
\mathbf{C}_{SS}\right) ^{-1}\mathbf{b}_{S}\left( \mathbf{G},\boldsymbol{%
\theta }\right),
\end{equation*}
which reduces to $\delta L_{ij}$ after some algebra.

For item (ii), we apply Proposition \ref{prop-1} with $S=\{i,j\}$, $\mathbf{C%
}_{SS}=-%
\begin{bmatrix}
0 & 1 \\ 
1 & 0%
\end{bmatrix}%
$. The analysis is similar, and hence omitted. \hfill $\Box$ \medskip 

\bibliographystyle{chicagoa}
\bibliography{bib-intervention}

\newpage

\begin{center} {\Large
Online Appendix \\
(Not for publication) }
\end{center}

\section{A walk-counting interpretation}

\label{app-bridge}

\subsection{Intercentrality of the group}

In this subsection, we show that any block of $\mathbf{W}\left( \mathbf{G}%
,S\right) $ can be decomposed by the idea of walk concatenations.We first
introduce some notation.

\begin{defin}
\label{def-P} A walk $\mathcal{P}$ in network $\mathbf{G}$ is a finite
sequence of nodes $(i_1, i_2,\cdots, i_{t+1})$ such that $%
g_{i_1i_2}=g_{i_2i_3}=\cdots=g_{i_ti_{t+1}}=1$. The node $i_1$ is the
starting node and $i_{t+1}$ is the ending node. The length of the path is
denoted by $\#(\mathcal{P})=t$. A path of length zero is supposed to be a
one-tuple $(i_1)$.

Fixing the parameter $\delta $, we define $\vee _{\delta }(\mathcal{P}%
):=\delta ^{\#(\mathcal{P})}$ for a walk $\mathcal{P}$. This definition is
linearly extended to a set of walks $\mathbb{P}$:$\vee _{\delta }(\mathbb{P}%
):=\sum_{\mathcal{P}\in \mathbb{P}}\vee _{\delta }(\mathcal{P})=\sum_{%
\mathcal{P}\in \mathbb{P}}\delta ^{\#(\mathcal{P})}. $ (We often drop the
subscript $\delta $ in $\vee $ when the context is clear.)

Given two walks $\mathcal{P}=(i_{1},i_{2},\cdots ,i_{t+1})$, $\mathcal{Q}%
=(j_{1},j_{2},\cdots ,j_{s+1})$ with $i_{t+1}=j_{1}$, we construct the
concatenation of $\mathcal{P}$ and $\mathcal{Q}$ as $\mathcal{P}\odot 
\mathcal{Q}=(i_{1},i_{2},\cdots ,i_{t+1},j_{2},\cdots ,j_{s+1})$. Clearly,
we have 
\begin{equation}
\vee (\mathcal{P}\odot \mathcal{Q})=\vee (\mathcal{P})\cdot \vee (\mathcal{Q}%
)\mbox{~and~ }\#(\mathcal{P}\odot \mathcal{Q})=\#(\mathcal{P})+\#(\mathcal{P}%
).
\end{equation}
\end{defin}

These definitions are convenient. For instance, let $\mathbb{M}_{ij}(\mathbf{%
G})$ denote the set of walks that starts with $i$ and ends at $j$ in
network $\mathbf{G}$. Then, we have $m_{ij}(\mathbf{G})=\vee (\mathbb{M}%
_{ij}(\mathbf{G}))$.

\begin{defin}
\label{def-W2} Fixing a non-empty proper subset $S$ of $N$ in the network $%
\left( N,\mathbf{G}\right) $, for any $i$, $j\in N$, we define $\mathbb{W}%
_{ij}\left( \mathbf{G},S\right) $ as the set of walks in $\mathbb{M}_{ij}(%
\mathbf{G})$ that does not contain any node in $S$ with the possible exception
of the starting node $i$ and the ending node $j$.
\end{defin}

By the definition, it is obvious that $w_{ij}\left( \mathbf{G},S\right)
=\vee (\mathbb{W}_{ij}\left( \mathbf{G},S\right) )$. Now we are in 
position to show the equations in Proposition \ref{prop-3} using walk counting.

Given $i\in S^{C},j\in S$, any walk $\mathcal{Q}$ in $\mathbb{M}_{ij}$ can
be \emph{uniquely} decomposed as the concatenation of two walks, $\mathcal{P%
}^{il}$ and $\mathcal{P}^{lj}$ for some $l\in S$, where $l$ is the \emph{%
first} node along the walk $\mathcal{Q}$ that that node is in $S$. This walk 
$\mathcal{Q}$ contains at least one node in $S$, since the ending node $j$ is in 
$S$. By this definition of $l$, we have $\mathcal{P}^{il}\in \mathbb{W}_{il}(%
\mathbf{G},S)$. Clearly, $\mathcal{P}^{lj}\in \mathbb{M}_{lj}$. Furthermore,
such a decomposition is unique. Consequently, we have the following
decomposition:
\begin{equation*}
\mathbb{M}_{ij}=\bigcup_{l\in S}\bigcup_{\mathcal{P}^{il}\in \mathbb{W}_{il}(%
\mathbf{G},S)}\bigcup_{\mathcal{P}^{lj}\in \mathbb{M}_{lj}}\mathcal{P}%
^{il}\odot \mathcal{P}^{lj}.
\end{equation*}%
Taking $\vee $ on both sides and applying the properties of $\vee $ yields
the following linear equations:
\begin{equation*}
m_{ij}=\sum_{l\in S}w_{il}m_{lj}
\end{equation*}%
for any $i\in S^{C},j\in S$. In matrix form, we obtain that $\mathbf{M}%
_{S^{C}S}\left( \mathbf{G}\right) =\mathbf{W}_{S^{C}S}\left( \mathbf{G}%
,S\right) \left( \mathbf{M}_{SS}\left( \mathbf{G}\right) \right) $. Hence, $%
\mathbf{W}_{S^{C}S}\left( \mathbf{G},S\right) =\mathbf{M}_{S^{C}S}\left( 
\mathbf{G}\right) \left( \mathbf{M}_{SS}\left( \mathbf{G}\right) \right)
^{-1}$.

Now we turn to equation \eqref{eq: Lemma 1 extension 1}. It suffices to show the
following:
\begin{equation}
\mathbf{M}_{S^{C}S^{C}}\left( \mathbf{G}\right) =\mathbf{W}%
_{S^{C}S^{C}}\left( \mathbf{G},S\right) +\mathbf{W}_{S^{C}S}\left( \mathbf{G}%
,S\right) \mathbf{M}_{SS^{C}}\left( \mathbf{G}\right) ,  \label{eq-scsc}
\end{equation}%
which holds since for $i,j\in S^{C}$, the walks $\mathbb{M}_{ij}$ can be
decomposed into disjoint unions: 
\begin{equation*}
\mathbb{M}_{ij}=\mathbb{W}_{ij}\bigcup \left( \bigcup_{l\in S}\bigcup_{%
\mathbb{P}^{il}\in \mathbb{W}_{il}}\bigcup_{\mathbb{P}^{lj}\in \mathbb{M}%
_{lj}}\mathbb{P}^{il}\odot \mathbb{P}^{lj}\right) .
\end{equation*}%
The intuition follows from a simple counting exercise. For a walk $\mathcal{Q%
}$ in $\mathbb{M}_{ij}$, either it does not contain any node in $S$ or it
contains at least one node in $S$. The set of walks in the former case is
precisely $\mathbb{W}_{ij}$, while the set of walks in the latter case is
precisely $\left( \bigcup_{l\in S}\bigcup_{\mathbb{P}^{il}\in \mathbb{W}%
_{il}}\bigcup_{\mathbb{P}^{lj}\in \mathbb{M}_{lj}}\mathbb{P}^{il}\odot 
\mathbb{P}^{lj}\right) $ (note that $l$ is the first node in a walk $%
\mathcal{Q}$ such that this node is in $S$). Taking operator $\vee $ on both
sides yields equation \eqref{eq-scsc}.

To show equation \eqref{eq: Lemma 1 extension 3}, it suffices to show that 
\begin{equation}
{{\mathbf{M}}_{SS}}({\mathbf{G}})=\underbrace{{{\mathbf{W}}_{SS}}({\mathbf{G}%
},S)}_{{\text{Walks never passing }}S}+\underbrace{\left( {{{\mathbf{W}}_{SS}%
}({\mathbf{G}},S)-{\mathbf{I}}}\right) \left( {{{\mathbf{M}}_{SS}}({\mathbf{G%
}})-{\mathbf{I}}}\right) }_{{\text{Walks always passing }}S},
\label{ss} \\
\end{equation}%
which follows from the following decomposition:\footnote{%
The intuition for the decomposition is similar, and hence omitted. Here $k$ is
the first node of a walk in $\mathbb{M}_{ij}(\mathbf{G})\backslash \mathbb{W}%
_{ij}(\mathbf{G},S)$ so that that node is in $S$. We must remove the walk
with length zero in $\mathbb{W}_{ik}$ and $\mathbb{M}_{kj}$ to guarantee the
uniqueness of concatenation decomposition, which explains the identity
matrix in equation \eqref{ss}.} 
\begin{equation*}
\mathbb{M}_{ij}(\mathbf{G})\backslash \mathbb{W}_{ij}(\mathbf{G}%
,S)=\bigcup_{k\in S}\bigcup_{\mathcal{P}^{ik}\in \mathbb{W}_{ik}(\mathbf{G}%
,S)\backslash \{(i)\}}\bigcup_{\mathcal{P}^{kj}\in \mathbb{M}_{kj}(\mathbf{G}%
)\backslash \{(k)\}}\mathcal{P}^{ik}\odot \mathcal{P}^{kj}
\end{equation*}%
for $i,j\in S$. Taking $\vee $ on both sides yields 
\begin{equation*}
m_{ij}-w_{ij}=\sum_{k\in S}{[w_{ik}(\mathbf{G},S)-\mathbf{1}%
_{\{k=i\}}][m_{kj}(\mathbf{G})-\mathbf{1}_{\{k=j\}}]},
\end{equation*}%
which holds for any $i,j\in S$, thus is equivalent to equation \eqref{ss}.

\bigskip

\subsection{Key bridge index}

In this section we prove the following identity: 
\begin{equation}  \label{bridge-3part}
b\left( \mathbf{G+E}_{ij}\right)-b\left( \mathbf{G}\right) =\delta
L_{ij}\left( \mathbf{N}^{1},\mathbf{N}^{2}\right) = \frac{\delta^2
m_{jj}\left( \mathbf{N}^{2}\right) b_{i}^{2}\left( \mathbf{N}^{1}\right)
+\delta^2 m_{ii}\left( \mathbf{N}^{1}\right) b_{j}^{2}\left( \mathbf{N}%
^{2}\right) +2\delta b_{j}\left( \mathbf{N}^{2}\right) b_{i}\left( \mathbf{N}%
^{1}\right) }{1-\delta ^{2}m_{jj}\left( \mathbf{N}^{2}\right) m_{ii}\left( 
\mathbf{N}^{1}\right) }
\end{equation}
$\forall i\in N_1, j\in N_2,$ where $\mathbf{G}$ is in the union of two
isolated networks. For ease of notation we drop the network variable in $%
m_{ii}, b_i, b_j, m_{jj}$.

Define $\mathbb{W}_{.i}=\bigcup_{i\in N} \mathbb{W}_{ji}$ as the set of
walks ending at $i$. Define $\mathbb{W}_{i.}=\bigcup_{l\in N} \mathbb{W}%
_{il} $ as the set of walks starting with $i$ Clearly, $\vee(\mathbb{W}%
_{.i})=\vee(\mathbb{W}_{i.})=\sum_{j\in N}m_{ij}=\sum_{l\in N}m_{il}=b_i$.
Note that the term on the left-hand side of equation \eqref{bridge-3part} is the
discounted sum of additional walks due to the new link $(i,j)$. We will
divide these new walks into four types and match each type to a term
on the right-hand side of equation \eqref{bridge-3part}.

\noindent(Type I) New walks starting at a node in $\mathbf{N}^1$ and ending
at a node in $\mathbf{N}^2$.

Take such a walk $\mathcal{Q}$. Suppose it passes the bridge link $(i,j)$
only once. It can be uniquely written as $\mathcal{P}^{.i}\odot (i,j)\odot 
\mathcal{P}^{j.}$, where $\mathcal{P}^{.i}\in \mathbb{W}_{.i}(\mathbf{N}^1)$
and $\mathcal{P}^{j.}\in\mathbb{W}_{j.}(\mathbf{N}^2)$ (see Figure \ref{fg:
bridge 1} for such an example). We have 
\begin{equation*}
\vee \left(\bigcup_{\mathcal{P}^{.i}\in \mathbb{W}_{.i}(\mathbf{N}%
^1)}\bigcup_{\mathcal{P}^{j.}\in\mathbb{W}_{j.}(\mathbf{N}^2)} \mathcal{P}%
^{.i}\odot (i,j)\odot \mathcal{P}^{j.} \right) =\vee \left( \mathbb{W}%
_{.i}\right) \times \delta \times \vee \left(\mathbb{W}_{j.}\right) =\delta
b_i b_j.
\end{equation*}

\begin{figure}[h]
\centering
\par
\includegraphics[scale=0.25]{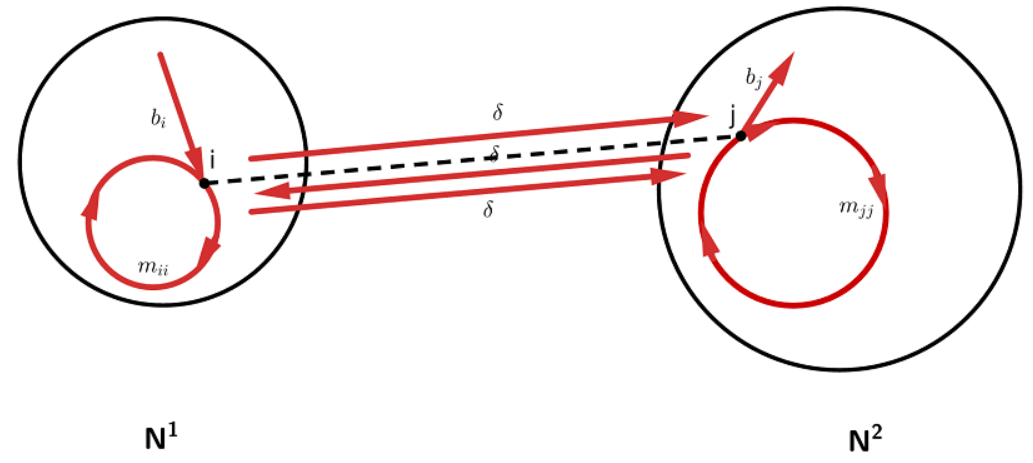}
\caption{New walk from one network to another by passing bridge three times}
\label{fg: bridge 2}
\end{figure}

However, $\mathcal{Q}$ can also pass the bridge link $(i,j)$ three times,
five times, etc. We can apply a similar exercise to show that the set of walks
originating from nodes in $\mathbf{N}^{1}$ and stopping at nodes in $\mathbf{%
N}^{2}$ that pass $\left( i,j\right) $ exactly three times is given by 
\begin{equation*}
\bigcup_{\left( k,l\right) \in N_{1}\times N_{2}}\bigcup_{(\mathcal{P}^{ki},%
\mathcal{P}^{jj},\mathcal{P}^{ii},\mathcal{P}^{jl})\in (\mathbb{M}_{ki}(%
\mathbf{N}^{1}))\times (\mathbb{M}_{jj}(\mathbf{N}^{2}))\times (\mathbb{M}%
_{ii}(\mathbf{N}^{1}))\times (\mathbb{M}_{jl}(\mathbf{N}^{2}))}(\mathcal{P}%
^{ki}\odot \left( i,j\right) \odot \mathcal{P}^{jj}\odot \left( j.i\right)
\odot \mathcal{P}^{ii}\odot \left( i,j\right) \odot \mathcal{P}^{jl})
\end{equation*}%
(Figure \ref{fg: bridge 2} gives a walk that passes the bridge three times.)
Taking $\vee$ yields 
\begin{eqnarray*}
&&\sum_{k\in N_{1}}\vee \left( \mathbb{M}_{ki}(\mathbf{N}^{1})\right) \cdot
\delta \cdot \vee \left( \mathbb{M}_{jj}(\mathbf{N}^{2})\right) \cdot \delta
\cdot \vee \left( \mathbb{M}_{ii}(\mathbf{N}^{1})\right) \cdot \delta \cdot
\sum_{l\in N_{2}}\vee \left( \mathbb{M}_{jl}(\mathbf{N}^{2})\right) \\
&=& \delta b_i b_j ( \delta^2 m_{ii}m_{jj}).
\end{eqnarray*}%
Similarly, $\delta b_i b_j ( \delta^2 m_{ii}m_{jj})^2$ captures Type I walks
that pass the link $(i,j)$ five times. Taking the sum yields 
\begin{equation}
\delta b_i b_j + \delta b_i b_j ( \delta^2 m_{ii}m_{jj}) + \delta b_i b_j (
\delta^2 m_{ii}m_{jj})^2+\cdots= \delta b_i b_j \frac{1}{1-\delta^2
m_{ii}m_{jj}}.
\end{equation}

\begin{figure}[h]
\centering
\par
\includegraphics[scale=0.25]{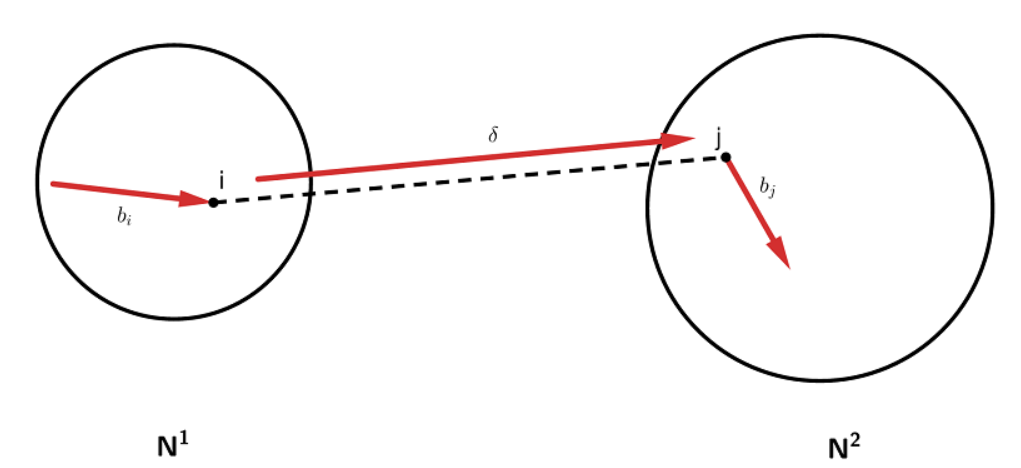}
\caption{New walk from one network to another by passing the bridge once}
\label{fg: bridge 1}
\end{figure}

\noindent(Type II) For the same reason, the new walks starting at a node in $%
\mathbf{N}^2$ and ending at a node in $\mathbf{N}^1$ contribute $\delta b_i
b_j \frac{1}{1-\delta^2 m_{ii}m_{jj}}$ to equation \eqref{bridge-3part}.

\begin{figure}[h]
\centering
\par
\includegraphics[scale=0.25]{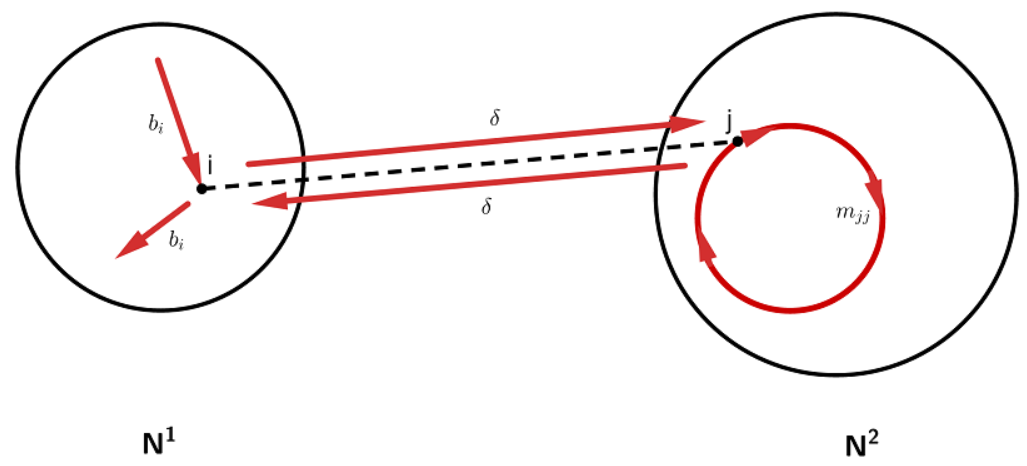}
\caption{New walk starting and ending at a same network}
\label{fg: bridge 3}
\end{figure}

\noindent(Type III) New walks starting at a node in $\mathbf{N}^1$ and
ending at a node in $\mathbf{N}^1$. By definition, such a walk must pass the
bridge $(i,j)$ two times, four times, etc. To pass the bridge twice, the
walk must be decomposed into the concatenation of the following walks: $%
\mathcal{P}^{.i}\odot \left( i,j\right) \odot \mathcal{P}^{jj}\odot \left(
j,i\right) \odot \mathcal{P}^{i.}$ where $\mathcal{P}^{.i}\in\mathbb{W}%
_{.i}, \mathcal{P}^{jj}\in\mathbb{W}_{ii} \mathcal{P}^{i.}\in\mathbb{W}_{.i}$
(see Figure \ref{fg: bridge 3} for an example). Taking $\vee$ of these walks
yields $\delta^2 b_i^2 m_{jj}$. To take into account the fact that such
walks can pass the bridge four times, six times, etc., we should multiply $%
\delta^2 b_i^2 m_{jj}$ by $\frac{1}{1-\delta^2 m_{ii}m_{jj}}$ (the
underlying logic is similar to the exercise for Type I). Together, Type III
walks contribute exactly $\delta^2 b_i^2 m_{jj}\frac{1}{1-\delta^2
m_{ii}m_{jj}}$ to equation \eqref{bridge-3part}.

\noindent(Type IV) New walks starting  at a node in $\mathbf{N}^2$ and
ending at a node in $\mathbf{N}^2$. This part is similar to Type III and 
contributes $\delta^2 b_j^2 m_{ii}\frac{1}{1-\delta^2 m_{ii}m_{jj}}$ to  equation
\eqref{bridge-3part}.

\section{Interventions in alternative network models}

\label{app-alt-C}

Our paper is applicable to many other network models in which 
Katz-Bonacich centrality plays an important role in shaping the equilibrium.
We list several models. A common theme is that  Katz-Bonacich centrality,
similar to \cite{Ballester2006}, is a building block of the equilibrium
objective.

\begin{enumerate}
\item \textbf{(Multiple activities)} \cite{D2x} consider a network model
with multiple interdependent activities. In the network $\left( N,\mathbf{G}%
\right) $, each player $i$ can choose the levels of two activities $\left(
a_{i}^{A},a_{i}^{B}\right) =\mathbf{a}_{i}$ with utility 
\begin{eqnarray*}
u_{i}\left( \mathbf{a}_{i},\mathbf{a}_{-i}\right) &=&\theta
_{i}^{A}a_{i}^{A}+\theta _{i}^{B}a_{i}^{B}-\left\{ \frac{1}{2}\left(
a_{i}^{A}\right) ^{2}+\frac{1}{2}\left( a_{i}^{B}\right) ^{2}+\beta
a_{i}^{A}a_{i}^{B}\right\} +\delta \underset{j}{\sum }%
g_{ij}a_{i}^{A}a_{j}^{A}+\delta \underset{j}{\sum }g_{ij}a_{i}^{B}a_{j}^{B}%
\text{,}
\end{eqnarray*}%
where $\boldsymbol{\theta }_{i}=\left( \theta _{i}^{A},\theta
_{i}^{B}\right) $ is $i$'s characteristics; $\frac{1}{2}\left(
a_{i}^{A}\right) ^{2}+\frac{1}{2}\left( a_{i}^{B}\right) ^{2}+\beta
a_{i}^{A}a_{i}^{B}$ is the cost of action $\mathbf{a}_{i}$;
 and $\delta \underset{j}{\sum }g_{ij}a_{i}^{A}a_{j}^{A}+\delta \underset{j}{\sum }%
g_{ij}a_{i}^{B}a_{j}^{B}$ captures the network externalities. \cite{D2x}
show that if $0\leq \delta \leq \frac{1-|\beta |}{\lambda _{\max }\left( 
\mathbf{G}\right) }$, there exists a unique Nash equilibrium given by%
\begin{equation*}
\begin{bmatrix}
\mathbf{x}^{A} \\ 
\mathbf{x}^{B}%
\end{bmatrix}%
=%
\begin{bmatrix}
\frac{1}{2\left( 1+\beta \right) }\mathbf{b}\left( \mathbf{G},\boldsymbol{%
\theta }^{A}+\boldsymbol{\theta }^{B},\frac{\delta }{1+\beta }\right) +\frac{%
1}{2\left( 1-\beta \right) }\mathbf{b}\left( \mathbf{G},\boldsymbol{\theta }%
^{A}-\boldsymbol{\theta }^{B},\frac{\delta }{1-\beta }\right) \\ 
\frac{1}{2\left( 1+\beta \right) }\mathbf{b}\left( \mathbf{G},\boldsymbol{%
\theta }^{A}+\boldsymbol{\theta }^{B},\frac{\delta }{1+\beta }\right) -\frac{%
1}{2\left( 1-\beta \right) }\mathbf{b}\left( \mathbf{G},\boldsymbol{\theta }%
^{A}-\boldsymbol{\theta }^{B},\frac{\delta }{1-\beta }\right)%
\end{bmatrix}%
\text{,}
\end{equation*}%
where $\mathbf{b}\left( \mathbf{G},\boldsymbol{\theta },\frac{\delta }{%
1+\beta }\right) =\left( \mathbf{I}-\frac{\delta }{1+\beta }\mathbf{G}%
\right) ^{-1}\boldsymbol{\theta }$. That is, the equilibrium profile is the
weighted sum of two Katz-Bonacich centralities.

\item \textbf{(Direct complements and indirect substitutes)}

\cite{CURRARINI201740} introduce a linear quadratic network game in which
each agent faces peer effects from  distance-one neighbors but also
exhibits local congestion effects from distance-two neighbors. Specifically,
the payoff of individual $i$ is given by%
\begin{equation*}
u_{i}\left( a_{i},\mathbf{a}_{-i}\right) =\theta _{i}a_{i}-\frac{1}{2}%
a_{i}^{2}+\delta \sum _{k=1}^{n}g_{ik}a_{i}a_{k}-\gamma \sum
_{k=1}^{n}g_{ik}^{\left[ 2\right] }a_{i}a_{k}\text{,}
\end{equation*}%
where the last term $-\gamma \Sigma _{k=1}^{n}g_{ik}^{\left[ 2\right]
}a_{i}a_{k}>0$ is new compared with \cite{Ballester2006} and captures the
strategic substitution effect between players at distance-two in the
network. Note that $\gamma\geq 0$ and $g_{ik}^{\left[ 2\right] }$ is the $ik$%
-th element of matrix $\mathbf{G}^{2}$. Under some regularity assumptions on 
$\delta $ and $\gamma $, \cite{CURRARINI201740} show that the unique
equilibrium equals 
\begin{equation*}
\mathbf{x}^{\ast }=\left( \mathbf{I}-\delta \mathbf{G}+\gamma \mathbf{G}%
^{2}\right) ^{-1}\boldsymbol{\theta }\text{.}
\end{equation*}%
In fact, we can rewrite the above equilibrium succinctly as a linear
combination of two Katz-Bonacich centralities: 
\begin{equation}  \label{eq:localcongest}
\mathbf{x}^{\ast }=\frac{\beta _{1}}{\beta _{1}-\beta _{2}}\mathbf{b}\left( 
\mathbf{G},\boldsymbol{\theta ,}\beta _{1}\right) -\frac{\beta _{2}}{\beta
_{1}-\beta _{2}}\mathbf{b}\left( \mathbf{G},\boldsymbol{\theta ,}\beta
_{2}\right) \text{.}
\end{equation}%
where $\beta _{1}=\frac{\delta +\sqrt{\delta ^{2}-4\gamma }}{2}$ and $\beta
_{2}=\frac{\delta -\sqrt{\delta ^{2}-4\gamma }}{2}$. Note that $\beta_1$ and 
$\beta_2$ satisfy $\beta_1+\beta_2=\delta, \beta_1 \beta_2=\gamma$. Equation
\eqref{eq:localcongest} directly follows from the following mathematical
identity: 
\begin{eqnarray*}
\left( \mathbf{I}-\delta \mathbf{G}+\gamma \mathbf{G}^{2}\right) ^{-1} &%
\boldsymbol{=}&\left( \mathbf{I}-\left( \beta _{1}+\beta _{2}\right) \mathbf{%
G}+\beta _{1}\beta _{2}\mathbf{G}^{2}\right) ^{-1} =\frac{\beta _{1}}{\beta
_{1}-\beta _{2}}\left( \mathbf{I-}\beta _{1}\mathbf{G}\right) ^{-1}-\frac{%
\beta _{2}}{\beta_{1}-\beta_{2}}\left( \mathbf{I-}\beta _{2}\mathbf{G}%
\right) ^{-1}\text{.}
\end{eqnarray*}
That is, the equilibrium in \cite{CURRARINI201740} equals the weighted sum
of two Katz-Bonacich centralities.

\item (\textbf{Local complementarity and global substitution)} In addition
to local complementaries, players might experience global competitive
effects. In one extension, \cite{Ballester2006} consider the following
utility function of individual $i$:
\begin{equation*}
u_{i}\left( a_{i},\mathbf{a}_{-i}\right) =a_{i}-\frac{1}{2}a_{i}^{2}-\phi
\Sigma _{k\neq i}a_{i}a_{k}+\delta \Sigma _{k=1}^{n}g_{ik}a_{i}a_{k}\text{,}
\end{equation*}%
where the term $\phi \Sigma _{k\neq i}a_{i}a_{k}$ is the global interaction
effect that corresponds to a substitutability in efforts across all players. $%
\phi \geq 0$ measures the intensity of the global interdependence. For
simplicty, we assume that each player's intrinsic marginal utilities are
identical. \cite{Ballester2006} show that the equilibrium behavior in this
network game is%
\begin{equation*}
\mathbf{x}=\frac{1}{1-\phi +\phi b\left( \mathbf{G},\frac{\delta }{1-\phi }%
\right) }\mathbf{b}\left( \mathbf{G},\frac{\delta }{1-\phi }\right) \text{.}
\end{equation*}%
Each player's equilibrium strategy is a function of Katz-Bonacich
centralities. In particular, the aggregate equilibrium action, $\mathbf{1}%
^{\prime }\mathbf{x=}\frac{b\left( \mathbf{G},\frac{\delta }{1-\phi }\right) 
}{1-\phi +\phi b\left( \mathbf{G},\frac{\delta }{1-\phi }\right) }$, is a
monotonic function of $b\left(\mathbf{G},\frac{\delta }{1-\phi }\right) $.
\end{enumerate}

Since the equilibrium in each of these models is either linear combinations
or transformations of the Katz-Bonacich centralities, we can directly apply
Proposition \ref{prop-1} to study the effects of structural and
chracteristical interventions, and study similar issues such as the key
group and the key link problems.


\end{document}